\documentclass[]{emulateapj}
\usepackage[T1]{fontenc}
\usepackage[latin9]{inputenc}
\usepackage{verbatim}
\usepackage{amssymb}

\usepackage{rotating}

\begin{document}

%\title{Secular Orbital Evolution in ``hot-Jupiter'' Systems}  
\title{SECULAR ORBITAL EVOLUTION OF COMPACT PLANET SYSTEMS} 

\author{Ke Zhang\altaffilmark{1}}
\author{Douglas P. Hamilton\altaffilmark{1}}
\author{Soko Matsumura\altaffilmark{1,2}}
\altaffiltext{1}{Department of Astronomy, University of Maryland, College Park, MD 20742, USA}
\altaffiltext{2}{School of Engineering, Physics, and Mathematics, University of Dundee, Scotland DD1 4HN}
%\affil{Department of Astronomy, University of Maryland}
%\affil{College Park, MD 20742, USA}
\email{Corresponding Authors: dphamil@umd.edu, soko@astro.umd.edu}

\begin{abstract}
%  Since some of the single, close-in planets are unable to maintain an
%  eccentric orbit against tidal circularization for more than several
%  hundred million years, we study the effects of a possible companion
%  and investigate the secular orbital evolution of a planar two-planet
%  system subject to eccentricity damping.
%  Although an eccentric orbit of a single, close-in planet is typically circularized 
%  within several hundred million years due to the tidal interactions with the central 
%  star, at least some close-in exoplanets seem to maintain eccentric orbits.  
  Recent observations have shown that at least some close-in
  exoplanets maintain eccentric orbits despite tidal circularization
  timescales that are typically much shorter than stellar ages.
% the tidal interactions between the planet and the star typically circularize such 
% an orbit within several hundred million years.   
%  Previous studies have shown that eccentricity damping timescales can be prolonged 
%  for gravitatioinally-interacting two-planet systems. 
%  Motivated by these results, 
  We explore gravitational interactions with a more distant planetary
  companion as a possible cause of these unexpected non-zero
  eccentricities.  For simplicity, we focus on the evolution of a
  planar two-planet system subject to slow eccentricity damping and
  provide an intuitive interpretation of the resulting long-term
  orbital evolution.
%We further apply our model to the observed exoplanetary systems.
  We show that dissipation shifts the two normal eigenmode frequencies and
  eccentricity ratios of the standard secular theory slightly, and we confirm 
  that each mode decays at its own rate. Tidal damping of the
  eccentricities drives orbits to transition relatively quickly
  between periods of pericenter circulation and libration, and
%  Solving
%  for these rates analytically, we find that the orbits evolve between 
%  periods of pericenter circulation and libration because, typically, one mode damps
%  significantly more slowly than the other.  
%  As a result, the orbits shift between periods of pericenter circulation and libration.
  the planetary system settles into a locked state in which the pericenters are
  nearly aligned or nearly anti-aligned.
%We test our analytical results with
%  both numerical integrations of the first-order secular equations and
%  direct N-body simulations, and find close agreement.

  Once in the locked state, the eccentricities of the two orbits
  decrease very slowly because of tides rather than at the much more rapid
  single-planet rate, and thus eccentric orbits, even for close-in
  planets, can often survive much longer than the age of the system.
%  This provides a possible explanation for at least some of the
%  non-zero eccentricities of close-in planets, assuming that they have
%  companions in more eccentric orbits.  
  Assuming that an observed close-in planet on an elliptical orbit is
  apsidally-locked to a more distant, and perhaps unseen companion, we
  provide a constraint on the mass, semi-major axis, and eccentricity
  of the companion. We find that the observed two-planet system
  HAT-P-13 might be in just such an apsidally locked state, with
  parameters that obey our constraint reasonably well. We also survey
  close-in single planets, some with and some without an indication of
  an outer companion. None of the dozen systems that we investigate
  provides compelling evidence for unseen companions. Instead, we
  suspect that (1) orbits are in fact circular, (2) tidal damping rates
  are much slower than we have assumed, or (3) a recent event has
  excited these eccentricities. Our method should prove useful for
  interpreting the results of both current and future planet searches.

\end{abstract}

%\keywords{Secular evolution; Extrasolar planetary systems; Orbital dynamics.}

\section{Introduction}  \label{sec.intro}
%
%Since the first Jupiter-sized planet was discovered outside the Solar
%System around 51 Pegasi in 1995 \citep{mayor1995355},
%nearly 800 extrasolar planets have been found orbiting Sun-like stars
%(e.g., the Extrasolar Planet Encyclopedia, {\sl
%  http://exoplanet.eu/}).  These extrasolar planetary systems are
%largely unlike our own Solar System.  The majority of the initially
%detected planets are Jupiter-sized, are located much closer to their
%stars, and display substantially larger eccentricities, compared to
%gas giants in the Solar System.  This gas-giant
%dominant picture has been rapidly changing as higher precision
%radial-velocity searches, as well as powerful transiting missions like
%Kepler and CoRoT, have discovered a larger number of less massive,
%smaller planets
%\citep[e.g.][]{howard2010653,mayor201111092497,batalha201212025852}.
%However, these new discoveries only seem to deepen the mystery of
%various architectures of planetary systems.

In the past decade, many mechanisms have been proposed to explain a
wide range of eccentricities seen among exoplanets including
planet-planet scattering, planet-disk interaction, mean-motion
resonance passage, and Kozai resonances (see, e.g., a review by
\citet{namouni2007233} and references there-in). All of these mechanisms
can excite orbital eccentricities effectively.  While the orbits of
long-period planets could stay eccentric for billions of years, those
of most close-in planets are likely to be circularized within stellar
ages because of tidal interactions between the stars and the planets,
except in some interesting special cases considered by
\cite{correia2012L23}.
%For planets with large semi-major axes, the excited eccentricities may last for
%billions of years and their orbits are still eccentric today.  Problem
%arises, however, for planets very close to stars with orbital periods
%typically less than a week, because their orbits are likely to be
%circularized by now due to tidal dissipation.
%\sedit{Recent developments regarding Kepler, RV searches...}

\begin{figure}%[t]
%  \centering\includegraphics[width=0.6\textwidth]{plots/per_ecc.eps}
  \plotone{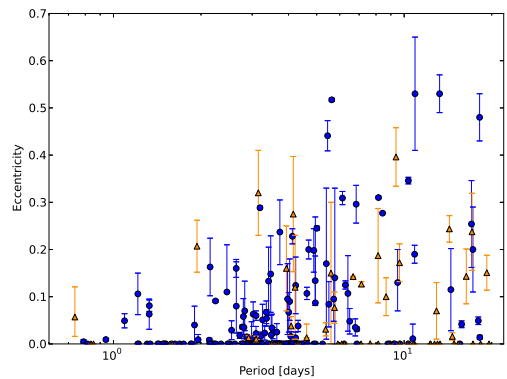}
%  \plotone{per_ecc.eps}
  \caption[Eccentricities of ``hot-Jupiters'']{Eccentricities of
    close-in planets with orbital periods less than 20 days. The blue
    circles and orange triangles correspond to single- and
    multiple-planet systems, respectively. Error bars are plotted for
    both period and eccentricity.  However, errors in period are not
    apparent because they are very small.  Orbital data, courtesy of
    the Exoplanet Orbit Database ({\sl http://exoplanets.org/}).}
  \label{fig.eccdist}
\end{figure}

In Figure~\ref{fig.eccdist}, we plot the eccentricities of all 222 known
(as of 2012 February) close-in planets with orbital periods of less than
20 days.
%with orbital periods less than 20 days in Fig.~\ref{fig.eccdist}, 
 %where data is taken from the Exoplanet Orbit Database ({\sl http://exoplanets.org/}). 
The average eccentricity for these planets ($\sim0.06$) is
significantly smaller than that of all of the confirmed planets
($\sim0.18$), which indicates that the tidal interaction
is an effective eccentricity damping mechanism and that
circularization timescales are typically shorter than stellar ages
(about 1-10\,Gyr).
%the figure indicates that many
%close-in systems indeed have short orbital circularization timescales
%compared to their stellar ages.  
Nevertheless, nearly half of these close-in planets have non-zero
orbital eccentricities.  Recent studies have shown, however, that the
orbital fits tend to overestimate eccentricities
\citep[e.g.,][]{shen2008553,zakamska20111895,pont20111278}, so some
measured eccentric orbits for close-in planets might in fact be circular.  
However, it is unlikely that all close-in planets have
perfectly circular orbits and thus at least some of the non-zero
eccentricities require a physical explanation. The main possibilities
include (1) systems have only recently attained their orbital
configurations and eccentricities are damping quickly, (2) planetary
tidal quality factors are much larger than those of the giant planets
in our solar system and so eccentricities damp slowly, and (3)
eccentricity excitation caused by an exterior planet slows the orbital
circularization.  In this paper, we explore the third option and
investigate the gravitational interactions between a close-in planet
and a more distant companion.

In systems with more than one planet, the most significant orbit-orbit
interactions are often mean-motion resonances, which have been studied
in detail for the satellite systems of the giant planets \citep[see
  reviews by][]{greenberg1977157, peale1986159}.  Resonances in
extrasolar planetary systems have also received increased attention
\citep[e.g.,][]{chiang2003465, beauge20031124, lee2004517,
  ketchum201371, batygin20131}. Mean-motion resonance passages during
planetary migration can be effective in exciting orbital
eccentricities.  These resonance passages can be divergent, in which
case impulsive changes to the orbital elements result, or convergent,
in which case trapping into resonance usually occurs
\citep{hamilton1994221, zhang2007386, zhang2008267}. Although
resonance capture typically leads to excited eccentricities, we do not
consider this process further here.  Instead, we focus on the more
prosaic secular interactions which are capable of maintaining orbital
eccentricities for a greater variety of orbital configurations.
%\citep{zhang20031485, zhang20071}. 

Secular perturbations have been studied for centuries in the context
of the Solar System \citep[see, e.g.,][]{brouwer1950}.  With the
discovery of extrasolar multi-planet systems \citep{butler1999916},
applications for secular theory have expanded
rapidly~\citep[e.g.,][]{wu20021024,barnes2006478,adams2006992,
  mardling20071768, batygin200923, greenberg20118,laskar2012105}.
\citet{wu20021024} were the first to use secular interactions to
explain the non-zero eccentricity of a hot Jupiter.
%Taking account of tidal evolution, and additional orbital precessions
%due to tidal and rotational deformations of planets as well as general
%relativity (GR) effects, t
They showed that a close-in planet with a companion could
sustain a substantial orbital eccentricity even though tidal
damping was efficient.  
%\citet{barnes2006478} used the classical
%secular theory and found that the known multi-planet systems tended to
%be in the secular separatrix between apsidal circulation and
%libration.  \citet{adams2006992} did secular analysis on several
%systems by including the GR corrections, and found that the effects
%could be significant, depending on the architectures of the planetary
%systems.  Taking account of an orbital circularization and the
%relativistic effect, 
\citet{zhang20031485} and \cite{zhang20071} confirmed the results of
\citet{wu20021024} and obtained most of the results contained in
Section~2 of the present article.
\citet{mardling20071768} developed a detailed octopole-order secular theory to
study orbital evolution of close-in planets with companions and
also confirmed the results of \citet{wu20021024}. 
%She has shown, both analytically and
%numerically, that the circularization time can be significantly
%prolonged for two-planet systems.  \citet{batygin200923} utilized her
%model, and proposed that the large observed eccentricity of GJ\,436b
%could be maintained if the observed planet and an outer hypothetical
%planet had evolved to an apsidally aligned, quasi-fixed state.  More
%recently, \citet{greenberg20118} presented a secular model with
%orbital decay and circularization due to tides, and showed that the
%quasi-fixed point behavior was modified by migration.  The same
%authors also further applied the model on 55 Cancri
%\citep{vanlaerhoven2012215} and 61 Virginis
%\citep{greenberg2012429}. \citet{laskar2012105} extended a similar
%secular model for N-planet systems.
%
%%kzh, need more working
%These studies,
%however, focused on just a few exoplanet systems.  Currently, there
%are nearly 100 confirmed multiple-planet systems.  Moreover, Kepler
%mission detected at least 170 multiple planet candidate systems, and
%most of these candidates are estimated to be actual planets
%\citep{lissauer201212015424}.  Thus, it is important to check these
%systems individually for the effects of secular interactions, and
%study why some close-in planets stay eccentric.
%
\citet{laskar2012105} pointed out the importance of precession due to
tidal effects. If these well-studied secular interactions play an
important role in delaying eccentricity damping, we might expect
differences in the eccentricity distributions between single- and
multiple-planet systems.  There is no obvious difference between these
two groups of planets, however, as can be seen in
Figure~\ref{fig.eccdist}. Perhaps future observations might reveal such
a difference, but it is also plausible that many single close-in planets
are actually accompanied by unobserved companions that help maintain their
eccentricities against tidal dissipation.
%exists and whether some of the observed single planets are accompanied by companions.

In this paper, we revisit the problem of secular interactions with a
distant companion in maintaining the eccentricities of close-in
planets.  Our goals are to develop an intuitive interpretation of the
secular theory of a two-planet system and to test the model against
observed planetary systems.  In the next section, we present a linear Laplace-Lagrange model 
for secular orbital evolution during tidal dissipation, starting with
a review of secular orbital interactions in a stable non-dissipative
system consisting of a star and two planets.  We then add tidal
dissipation of eccentricities, and solve the coupled system to
investigate how eccentricity damping affects the apsidal state of the
two orbits. We add the precession due to planetary and
stellar tidal and rotational bulges as well as general relativity (GR) 
terms which can be significant for close-in planets. In
Section~\ref{sec.results}, we apply this model to extrasolar planetary
systems, illustrating how it might help to guide planet searches.
Last, we discuss and summarize our work in Section~\ref{sec.conclusions}.

%
%================================================================================================
% SECTION 2
\section{Model} \label{sec.model}
\subsection{Stable Non-dissipative Systems with Two Planets} \label{sec.stable}

The secular solution of a stable non-dissipative 2-planet system is
known as Laplace-Lagrange theory and is discussed in great detail in
\citet{murray1999ssd}.  In this section, we will develop a graphical
interpretation of the solution that will help us understand the more
complicated systems studied later in this paper.  After averaging the
disturbing functions of each planet on the other, $\mathcal{R}_j$,
over both planets' orbital periods, Lagrange's planetary equations can
be linearized for small eccentricities and inclinations to \citep[][Section~7.1]{murray1999ssd}
\begin{eqnarray}
&&\dot{a}_j = 0; \nonumber\\
&&\dot{e}_j = - \frac{1}{n_j a_j^2 e_j}
  \frac{\partial\mathcal{R}_j}{\partial \varpi_j},\;\;
\dot{\varpi}_j = + \frac{1}{n_j a_j^2 e_j}
  \frac{\partial\mathcal{R}_j}{\partial e_j}; \label{eq.lang}\\
&&\dot{I}_j =  - \frac{1}{n_j a_j^2 I_j}
  \frac{\partial\mathcal{R}_j}{\partial \Omega_j}, \;\;
\dot{\Omega}_j = + \frac{1}{n_j a_j^2 I_j}
  \frac{\partial\mathcal{R}_j}{\partial I_j}. \nonumber
\end{eqnarray}
Here, $n_j$, $a_j$, $e_j$, $I_j$, $\Omega_j$, and $\varpi_j$ are the
mean motion, semi-major axis, eccentricity, inclination, longitude of
ascending node, and argument of the pericenter of the $j^{\rm th}$
planetary orbit, respectively.

Planetary semi-major axes remain constant and hence no long-term
energy transfer between the orbits occurs because we have averaged a
conservative perturbation force over each planet's orbital period.
Note that because of our assumption of small eccentricities and
inclinations, the evolution equations of ($e_j$, $\varpi_j$) and
($I_j$, $\Omega_j$) are completely decoupled and can be analyzed
separately.  In this paper, we neglect the inclination and node pair
since they are less easily observable for extrasolar planets.  Because
the two sets of equations take the same form, however, our
eccentricity results below can be easily applied to secular coupling
of vertical motions.

The disturbing function $\mathcal{R}$ is defined for a planet as the
non-Keplerian potential at its location \citep[][Section~6]{murray1999ssd}.
For two-planet systems without any external perturbations, it is
simply the gravitational potential caused by the other planet.  Here, we
consider a simple planetary system consisting of a central star and
two planets in co-planar orbits.  In terms of osculating elements
\citep[see, e.g.,][Section~2.9]{murray1999ssd} and to second order in small
eccentricities, \citet{murray1999ssd} show that the orbit-averaged
disturbing functions are given by:
\begin{eqnarray}
  &&\mathcal{R}_1 = n_1 a_1^2 \, \sigma q \left[ \frac{1}{2}e_1^2 - \beta \,
    e_1e_2\cos(\varpi_1-\varpi_2)\right], \label{eq.R1} \\
  &&\mathcal{R}_2 = n_2 a_2^2 \, \sigma \sqrt{\alpha} \left[
    \frac{1}{2}e_2^2 - \beta \, e_1e_2\cos(\varpi_1-\varpi_2)
    \right], \label{eq.R2}
\end{eqnarray}
where the subscript ``1'' refers to the inner planet and ``2'' to the
outer one.  We define the mass ratio between the two planets $q =
m_2/m_1$, and the semi-major ratio of the two orbits $\alpha = a_1/a_2$.
The remaining parameters are defined as follows:
\[
\beta = \frac{b_{3/2}^{(2)}(\alpha)}{b_{3/2}^{(1)}(\alpha)} \,\mathrm{and~} \sigma=\frac{1}{4} n_1 \frac{m_1}{m_*}\alpha^2 
b_{3/2}^{(1)}(\alpha),
\]
where $m_*$ is the stellar mass and $b_{3/2}^{(1)}(\alpha)$ and
$b_{3/2}^{(2)}(\alpha)$ are two of the Laplace coefficients
\citep{murray1999ssd}. The parameter $\sigma$ has units of frequency,
which we will soon see characterizes the secular precession
rates. Both $\beta$ and $\sigma$ decrease with increasing planetary
separation and, for small $\alpha$, the Laplace coefficients reduce to
$b_{3/2}^{(1)}(\alpha) \approx 3\alpha$ and $b_{3/2}^{(2)}(\alpha)
\approx 15\alpha^2/4$.

Following \citet{brouwer1950}, we transform a ($e_j$, $\varpi_j$) pair
into a complex Poincar\'{e} canonical variable $h_j$ with the mapping:
\begin{equation}
h_j = e_j \, \exp(i \varpi_j), \label{eq.hdef}
\end{equation}
where $i=\sqrt{-1}$.  Substituting Equations~(\ref{eq.R1}) and (\ref{eq.R2}) into
Equation~(\ref{eq.lang}) and rewriting in terms of $h_j$ yields a set of
linear homogeneous ordinary differential equations similar to those
for a double pendulum system:
\begin{equation}
\dot{h}_j = i \sum_{k=1}^2 A_{jk} h_k,  \label{eq.hdot}
\end{equation}
where the coefficient matrix
\[
A= \sigma \left\{ \begin{array}{cc}
     q & - q \beta \\
     - \sqrt{\alpha} \beta & \sqrt{\alpha}
     \end{array} \right\}.
\]

Two orthogonal special solutions of Equation.~(\ref{eq.hdot}), or the secular
eigenmodes of the system, are given by 
\[
\left( \begin{array}{l} \hat{h}_{1\pm} \\ \hat{h}_{2\pm} \end{array} \right)=
\left( \begin{array}{l} 1 \\ \eta_{\pm}^s \end{array} \right) 
\exp(i g_{\pm}^st), 
\]
where the eigen-frequencies $g_{\pm}^s$ and eigenvector parameters
$\eta_{\pm}^s$ can be obtained from the matrix $A$:
\begin{eqnarray}
g_{\pm}^s &=& \frac{\sigma}{2} \left[ q+\sqrt{\alpha} \mp
  \sqrt{(q-\sqrt{\alpha})^2 + 4q\sqrt{\alpha} \beta^2} \right],
  \label{eq.gs} \\
\eta_{\pm}^s &=& \frac{q-\sqrt{\alpha} \pm \sqrt{(q-\sqrt{\alpha})^2 +
  4q\sqrt{\alpha} \beta^2}} {2q\beta}. \label{eq.etas}
\end{eqnarray}
We use the superscript ``$s$'' to indicate that those parameters are
for a ``static'', or non-dissipative, system. 
%We have chosen our 
The eigenmode components $\hat{h}_{1\pm}$ and $\hat{h}_{2\pm}$ 
depend only on the fixed constants $\alpha$ and $q$ through
Equations. (\ref{eq.gs}) and (\ref{eq.etas}), and not on the initial
eccentricities and pericenter angles.

\begin{figure}%[t]
%  \centering\includegraphics[width=0.6\textwidth]{plots/modes1.eps}
  \plotone{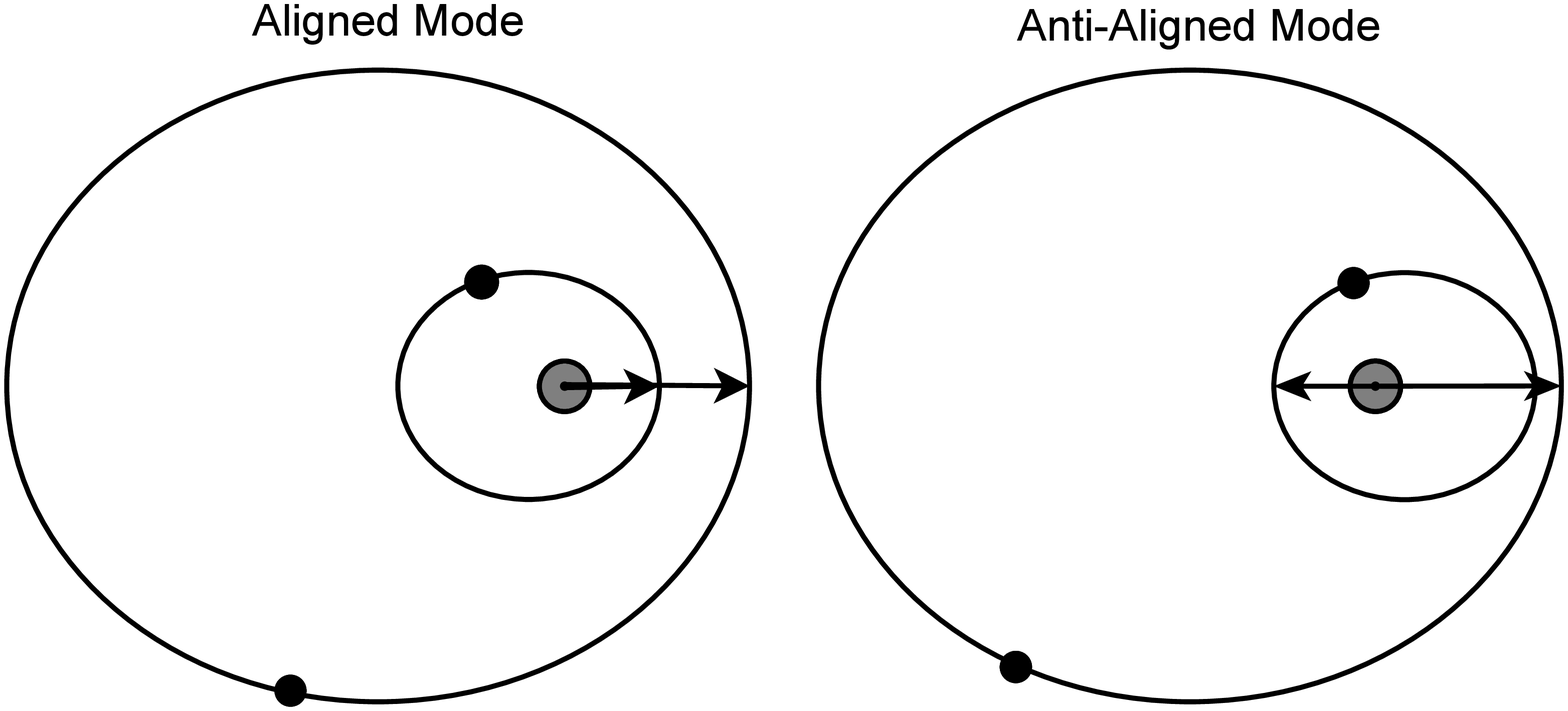}
  \caption{Aligned and anti-aligned secular modes. The planets (solid
    dots) follow nearly elliptical orbits about the central star.
    Arrows point from the star to orbital pericenters.}
  \label{fig.modes}
\end{figure}

The physical meaning of the two modes can be elucidated by
transforming the solution of $h_j$ back to the orbital elements ($e$,
$\varpi$) with Equation~(\ref{eq.hdef}).  If the system is fully in either
the ``+'' or the ``--'' mode, we find the following:
\begin{eqnarray}
&\dot{\varpi}_{1\pm}=\dot{\varpi}_{2\pm}=g_{\pm}^s, & \label{eq.rate} \\
& (e_2/e_1)_{\pm} = |\eta_{\pm}^s|,  &
  \label{eq.e2e1} \\
&\cos(\Delta\varpi_{\pm}) = \eta_{\pm}^s/|\eta_{\pm}^s|, &
  \label{eq.align} 
\end{eqnarray}
where $\Delta\varpi=\varpi_{2}-\varpi_{1}$ is the difference between
the two pericenter angles.  In either mode, the two orbits precess
together at the rate $g_{\pm}$ (Equation~(\ref{eq.rate})), and their
eccentricities keep a fixed ratio (Equation~(\ref{eq.e2e1})).  Furthermore,
Equation~(\ref{eq.etas}) shows that $\eta_+^s>0$ while $\eta_-^s<0$.  Thus,
Equation~(\ref{eq.align}) states that the pericenters of the two orbits are
always aligned in the ``+'' mode ($\cos(\Delta\varpi)=1$), and
anti-aligned in the ``--'' mode ($\cos(\Delta\varpi)=-1$).  
In an eigenmode, the system behaves as a rigid body with the shapes
and relative orientation of the elliptical orbits remaining fixed
(Figure~\ref{fig.modes}).  The frequency of the anti-aligned mode is
faster ($g_->g_+$) because of closer approaches between the planetary
orbits and thus stronger perturbations (Figure~\ref{fig.modes}).

% {\bf do we really need this?} I'd like to keep it and have
% simplified it somewhat
It is instructive to consider the small $q$ limit which corresponds to
a tiny outer mass. In this case, Equations~(\ref{eq.gs}) and (\ref{eq.etas})
simplify to ($g^s_+ = \sigma q (1-\beta^2)$, $\eta^s_+ = \beta$) and
($g^s_- = \sigma\sqrt\alpha$, $\eta^s_- = -\sqrt\alpha/q\beta$).  For
a true outer test particle $q\rightarrow 0$ and only the first mode is
possible. This aligned mode is stationary ($g_+=0$) and, since
$\eta_+^s < 1$, the planet's eccentricity exceeds that of the test
particle.  Similarly, with an inner test particle only the aligned
mode survives, it is stationary, and the massive planet again has the
higher eccentricity.

\begin{figure}%[t]
%  \centering\includegraphics[width=0.6\textwidth]{plots/eta.eps}
  \plotone{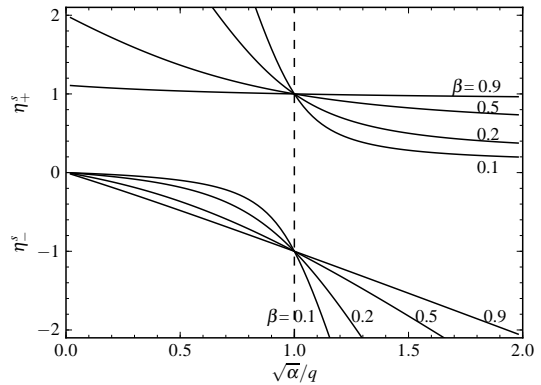}
  \caption{Eccentricity ratios vs. $\sqrt{\alpha}/q$ in the pure
    secular modes as given by Equation~(\ref{eq.etas}). Different curves
    represent different $\beta$ values.  Along the dashed line
    $q=\sqrt{\alpha}$ (or $m_1^2a_1 = m_2^2a_2$) and
    orbits have $e_1=e_2$ in either mode.}
  \label{fig.eta}
\end{figure}

Equation~(\ref{eq.e2e1}) gives the ratio between the eccentricities in a
perfect secular eigenmode. In Figure~\ref{fig.eta}, we plot
$\eta_{\pm}^s$ as a function of $\sqrt{\alpha}/q$ for different
$\beta$ values.  Although rare in real systems, $q=\sqrt{\alpha}$ (or
$m_1^2a_1=m_2^2a_2$) makes an interesting special case.  When this
condition is met, the planets have similar angular momenta and the
inner and outer planet orbital precession rates (the diagonal terms of
the matrix A) are equal. In addition, $\eta_{\pm}^s=\pm 1$
(Equation~(\ref{eq.etas})), and therefore the inner and outer orbits have the
same eccentricity (Equation~(\ref{eq.e2e1})).  When $q>\sqrt{\alpha}$
($m_1^2a_1 < m_2^2a_2$), the inner planet precesses fastest, has a
lower eccentricity in the aligned mode and a higher eccentricity in
the anti-aligned mode. The opposite is true for $q<\sqrt{\alpha}$.

\begin{figure*}%[p]
  \noindent\begin{tabular}{cc}
  \raisebox{-\height}{\includegraphics[width=0.48\textwidth]{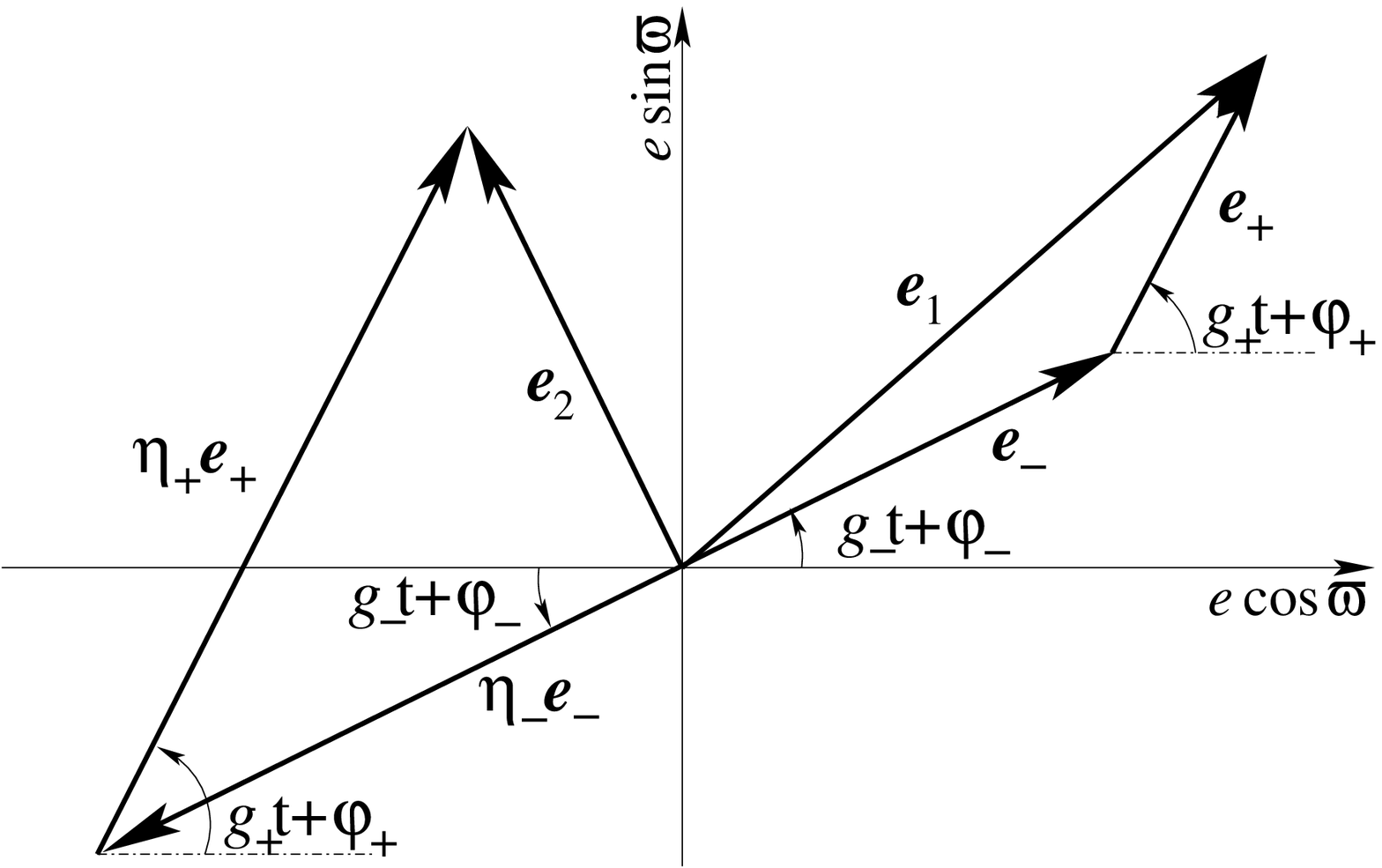}} &
  \raisebox{-\height}{\includegraphics[width=0.48\textwidth]{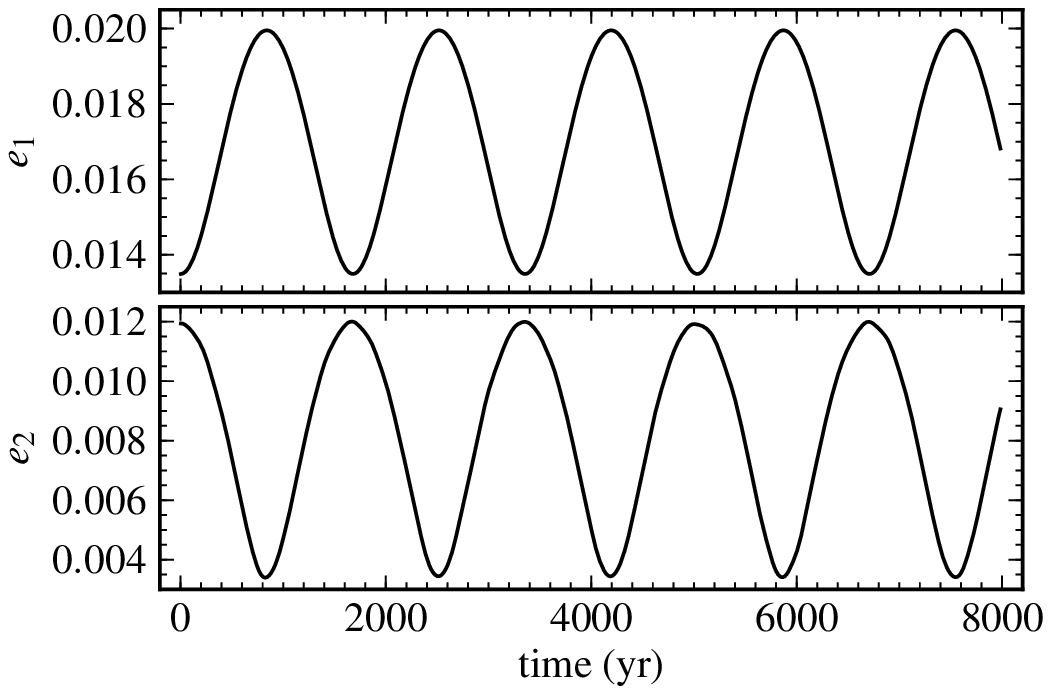}}\\
  (a) & (b)\\
  \end{tabular}
  \caption{General solution for a two-planet system.  (a) The solution
    (Equation~(\ref{eq.hsoln})) on a phase plot.  The arrows represent the
    eccentricity vector $e\exp(i\omega)$.  The total eccentricity of
    each orbit ($e_1$ or $e_2$) is the magnitude of the vector sum of aligned ($+$) and
    anti-aligned ($-$) components.  The two aligned components,
    $\mathbf{e}_+$ and $\eta_+^s\mathbf{e}_+$, rotate (precess) at
    rate $g_+^s$, while the two anti-aligned components
    ($\mathbf{e}_-$ and $\eta_-^s\mathbf{e}_-$) rotate at rate
    $g_-^s$. (b) Secular evolution of orbital eccentricities from an
    N-body simulation. Plot shows the eccentricities of two planetary
    orbits in a computer-simulated system consisting of a 1 Solar-mass
    star, a 1~Jupiter-mass hot-Jupiter at 0.05~AU, and a
    0.8~Jupiter-mass companion at 0.2~AU.  The simulation shows an
    oscillation period of $\sim 1673$ yr, in good agreement with
    the prediction of the secular model: $2\pi/(g_--g_+)=1670$ yr.}
  \label{fig.soln}
\end{figure*}

In general, a system is in a mixed state composed of a linear
combination of the two modes:
\begin{equation}
h_j=e_+\,\exp(i\varphi_+)\, \hat{h}_{j+} + e_- \, \exp(i\varphi_-) \,
\hat{h}_{j-}.  \label{eq.hsoln}
\end{equation}
Here the mode amplitudes $e_{\pm}$ and phases $\varphi_{\pm}$ are
determined by the initial orbits.  Plotting the solution
Equation~(\ref{eq.hsoln}) on the complex plane yields a phase plot of
$e\cos\varpi$ versus $e\sin\varpi$ as shown in Figure~\ref{fig.soln}a.
The $\mathbf{e}_1$ and $\mathbf{e}_2$ vectors in the plot represent
($e$, $\varpi$) pairs for the two orbits at a given time.  The length
of a vector is the instantaneous eccentricity, and its polar angle is
the instantaneous longitude of pericenter.  Each eccentricity vector
is a vector sum of an aligned component ($\mathbf{e}_+$ and
$\eta_+^s\mathbf{e}_+$ are parallel) and an anti-aligned component
($\mathbf{e}_-$ and $\eta_-^s\mathbf{e}_-$ are antiparallel).  The
lengths of all of these vectors are determined by initial conditions.
In a pure aligned eigenmode, $e_- = |\mathbf{e_-}|=0$ and the
$\mathbf{e}_1$ and $\mathbf{e}_2$ vectors are parallel.  As time
progresses, these eccentricity vectors rotate together at rate $g_+^s$
while maintaining their lengths.  This solution corresponds to the
aligned orbits in Figure~\ref{fig.modes}.  For the anti-aligned
eigenmode, $e_+ = |\mathbf{e_+}| =0$ so that $\mathbf{e}_1$ and
$\mathbf{e}_2$ are anti-parallel and rotate together at rate $g_-^s$.
This state is depicted by the pair of anti-aligned orbits in
Figure~\ref{fig.modes}.  In the most general system, both motions occur
simultaneously: the parallel eccentricity vectors ($\mathbf{e}_+$ and
$\eta_+^s\mathbf{e}_+$) rotate at rate $g_+^s$, while the
anti-parallel vectors ($\mathbf{e}_-$ and $\eta_-^s\mathbf{e}_-$)
rotate at rate $g_-^s$.  The resulting lengths of the eccentricities
$\mathbf{e}_1$ and $\mathbf{e}_2$, thus vary periodically as
illustrated in Figure~\ref{fig.soln}b.  The maximum value of $e_1$
occurs when $\mathbf{e}_+$ is parallel to $\mathbf{e}_-$.  At the same
time, however, $\eta_+^s\mathbf{e}_+$ and $\eta_-^s\mathbf{e}_-$ are
anti-parallel to each other, leading to a minimum value for $e_2$.
The simultaneous maximum for $e_1$ and minimum for $e_2$ seen in
Figure~\ref{fig.soln}b is a general result guaranteed by angular
momentum conservation.  The two eccentricities, in mathematical form
are as follows:
\begin{eqnarray}
&& e_1= \sqrt{e_+^2 + e_-^2 + 2e_+e_-\cos(g_-^s-g_+^s)t}, \nonumber \\
&& e_2= \sqrt{(e_+\eta_+^s)^2 + (e_-\eta_-^s)^2 +
  2e_+e_-\eta_+^s\eta_-^s\cos(g_-^s-g_+^s)t} \nonumber 
\end{eqnarray}
\citep{murray1999ssd}. Both eccentricities oscillate at the same frequency $(g_-^s-g_+^s)$ as
shown in Figure~\ref{fig.soln}b.

\subsection{Secular Modes with Eccentricity Damping} \label{sec.damping}

Since a planet close to its host star experiences a drag force caused by 
planetary tides raised by the star, we seek a way to include tides
into the mathematical formalism of the previous section.  For small
eccentricities for which the secular solution Equation~(\ref{eq.hsoln}) is
valid, tidal changes in $a$ are usually negligible compared with the
damping in $e$ \citep{goldreich1963257}.  Since the damping
timescales for close-in planets (in the order of $10^8$ yr) are much
longer than the secular timescales (typically $\sim 10^3$ yr), we
treat the damping effect as a small perturbation to the secular
solution.

Stars and planets raise tides on each other, which, in turn, perturb the
planet's orbit.  For very close-in planets, the combination of stellar
tides raised by the planet and planetary tides raised by the star
typically act to decrease the planet's orbital period and
eccentricity. Tidal dissipation in the star usually leads to orbital
decay and circularization on a timescale much longer than the age of
most planetary systems, while that tides in the planet can damp its orbital
eccentricity rather quickly
\citep[e.g.,][]{goldreich1963257,burns1977113,rasio19961187}.  Thus, we
assume that the orbital circularization is largely dominated by
planetary tides, and approximate the eccentricity damping rate as
follows \citep[e.g.,][]{murray1999ssd}:
\begin{equation}
\lambda=- \frac{\dot{e}}{e}=\frac{63}{4}\frac{1}{Q_p^{\prime}} 
  \frac{m_*}{m_p}\left(\frac{R_p}{a}\right)^5 n \, ,  \label{eq.erate}
\end{equation}
%
%where $m_*$ and $m_p$ are the masses of the star and the planet,
%respectively, $R_p$ is the planet's radius, $a$ is the orbital
%semi-major axis, and $n$ is the orbital mean motion.
where $Q_p^{\prime}\equiv1.5Q_p/k_2$ is the modified tidal quality factor, 
$Q_p$ is the tidal quality factor, and $k_2$ is the Love number of degree 2.  
%A recent work estimates $Q_p/k_2=(0.907\pm0.167)\times10^5$ for dissipation in
%Jupiter \citep{lainey2009957}, which corresponds to 
%$Q_p^{\prime}=(1.36\pm0.251)\times10^5$.
%
%We assume most extra-solar hot Jupiters and hot Neptunes have similar
%tidal dissipation efficiencies.  Since the eccentricity damping
%timescale is likely to be less than their stellar ages (typically
%1-10\,Gyr) for most of these systems,
%many close-in planets are expected to have circular orbits.

The addition of the constant tidal damping (Equation~(\ref{eq.erate})) adds
an extra term to the eccentricity equation in Equation~(\ref{eq.lang}),
which now reads as follows:
\[
\dot{e}_j  =  - \frac{1}{n_j a_j^2 e_j}
  \frac{\partial\mathcal{R}_j}{\partial \varpi_j} - \lambda_j e_j,
\]
with $\lambda_j$ being the eccentricity damping rate.
The coefficient matrix of Equation~(\ref{eq.hdot}) is now the following:
\begin{equation}
A= \sigma \left\{ \begin{array}{cc}
     q + i\,\xi_1 & - q \beta \\
     - \sqrt{\alpha} \beta & \sqrt{\alpha} + i\,\xi_2
     \end{array} \right\},  \label{eq.A}
\end{equation}
where the dimensionless $\xi_j=\lambda_j/\sigma \ll 1$ parameterizes the damping strength.

Eccentricity damping causes the eigen-frequencies of matrix $A$ to
have both real and imaginary parts.  As in other dynamical systems, the
real parts ($g_{\pm}$) of the eigen-frequencies still represent the
precession rates of the secular modes, while the imaginary parts
($\gamma_{\pm}$) indicate that the amplitudes of the modes change over
time.  This becomes clearer if we rewrite the two orthogonal special
solutions of the system as follows:
\begin{equation}
\left( \begin{array}{l} \hat{h}_{1\pm} \\ \hat{h}_{2\pm} \end{array} \right)=
\left( \begin{array}{l} 1 \\ \eta_{\pm} \end{array} \right) 
\exp(-\gamma_{\pm}t) \, \exp(i g_{\pm} t), 
\label{eq.hor}
\end{equation}
where $\eta_{\pm}$ is the new eccentricity ratio for each mode.

For $\xi_j \ll 1$, we solve the new matrix for the complex frequencies
and find the following: 
\begin{eqnarray}
g_{\pm} &=& g_{\pm}^s \pm \frac{q \sqrt{\alpha} \beta^2}
  {[(q - \sqrt{\alpha})^2 + 4\, q \sqrt{\alpha} \beta^2]^{3/2}} 
  \,\sigma \, (\xi_1-\xi_2)^2, \label{eq.g} \\
\gamma_{\pm} &=& \frac{1}{2} \left[ \lambda_1 + \lambda_2 \pm \frac{
    \sqrt{\alpha} - q} {\sqrt{(q - \sqrt{\alpha})^2 + 4\, q \sqrt{\alpha}
      \beta^2} } (\lambda_1-\lambda_2) \right].  \label{eq.gamma}
\end{eqnarray}
Eccentricity damping increases the precession rate of the aligned mode
and decreases that of the anti-aligned mode, but only by an extremely small
amount (of order $\xi_j^2$). These tiny frequency changes were neglected by
 \cite{laskar2012105}, but otherwise our damping rates are in perfect agreement.
The tiny frequency changes are due to the slightly different orbital
configurations in the secular modes as we shall describe below.  
The new eigen-vectors of the matrix $A$ are as follows:
\begin{equation}
  \eta_{\pm} = \eta_{\pm}^s \left[ 1 \pm i \frac{\xi_1-\xi_2}
    {\sqrt{(q - \sqrt{\alpha})^2 + 4\, q \sqrt{\alpha} \beta^2} }
    \right], \label{eq.eta}
\end{equation}
where we have neglected terms of second- and higher-order power of
$\xi_j$.  

In general, the $\eta_{\pm}$'s are complex with small imaginary
components.  If we ignore the imaginary parts for the moment, then
$\eta_{\pm}$ are real and Equation~(\ref{eq.hor}) shows that the pericenter
angles of the two orbits are the same (positive $\eta_+$, aligned
mode) or $180^{\circ}$ apart (negative $\eta_-$, anti-aligned mode).
For complex $\eta_{\pm}$, however, the two orbits are not exactly
aligned or anti-aligned any longer.  Instead,
$\Delta\varpi_{\pm}$ shifts from $0^{\circ}$ and $180^{\circ}$ by a
small angle
\small
\[
  \epsilon= \tan^{-1}\left(\frac{\xi_1-\xi_2}
    {\sqrt{(q - \sqrt{\alpha})^2 + 4\, q \sqrt{\alpha} \beta^2} }
  \right) \approx
  \frac{\xi_1-\xi_2}
    {\sqrt{(q - \sqrt{\alpha})^2 + 4\, q \sqrt{\alpha} \beta^2} }.
\]
\normalsize

In the ``aligned'' mode, the new pericenter difference is
$\Delta\varpi_+=\epsilon$ so that the inner exoplanet's pericenter
slightly lags that of the outer exoplanet. The lag is maximized for
$q=\sqrt\alpha$, the case with equal eccentricities and equal
precession rates for the two planets. Because of this mis-alignment,
the minimum distance between the two orbits is slightly less than that
in the undamped case (see Figure~\ref{fig.modes}). This causes the
average interaction between the two orbits to be stronger, leading to
an increase of the precession frequency as indicated by
Equation~(\ref{eq.g}).  Similarly, $\Delta\varpi_-=180^{\circ}-\epsilon$ in
the ``anti-aligned'' mode; the slight rotation results in a weaker
average interaction and a slightly slower mode-precession rate.  The
deviation angle $\epsilon$ is tiny, and the eccentricity ratios in the
two modes, $|\eta_{\pm}|$, are nearly the same as $|\eta_{\pm}^s|$.
Thus, we continue to use ``aligned'' and ``anti-aligned'' to refer to
the two modes. Nevertheless, the small deviation angle is physically
important because it is what enables monotonic damping of the outer
planet's eccentricity.

In addition to the slight mis-alignment, each mode amplitude also
damps at the rate given by Equation~(\ref{eq.gamma}).  If only planetary
tides contribute to eccentricity damping, Equation~(\ref{eq.erate}) shows
that the damping rate decreases rapidly with the planet's semi-major axis
($\lambda \propto a^{-6.5}$).  In the absence of secular interactions
between the planets, the outer orbit is hardly affected.  With this
interaction, however, the damping applied to the eccentricity of the
inner orbit is partially transmitted to the outer planet, causing
a decrease of its eccentricity as well.  The damping rates of the two
modes are different, unless $q=\sqrt{\alpha}$.
An interesting result from Equation~(\ref{eq.gamma}) is that the sum of the
two mode-damping rates is equal to the sum of the two individual
eccentricity damping rates:
\[
\gamma_+ + \gamma_-=\lambda_1 + \lambda_2.
\]

\noindent
The physical interpretation of this expression is that secular
interactions between the planets simply act to redistribute where the
damping occurs.

\begin{figure*}%[p]
  \noindent\begin{tabular}{cc}
    \includegraphics[width=0.48\textwidth]{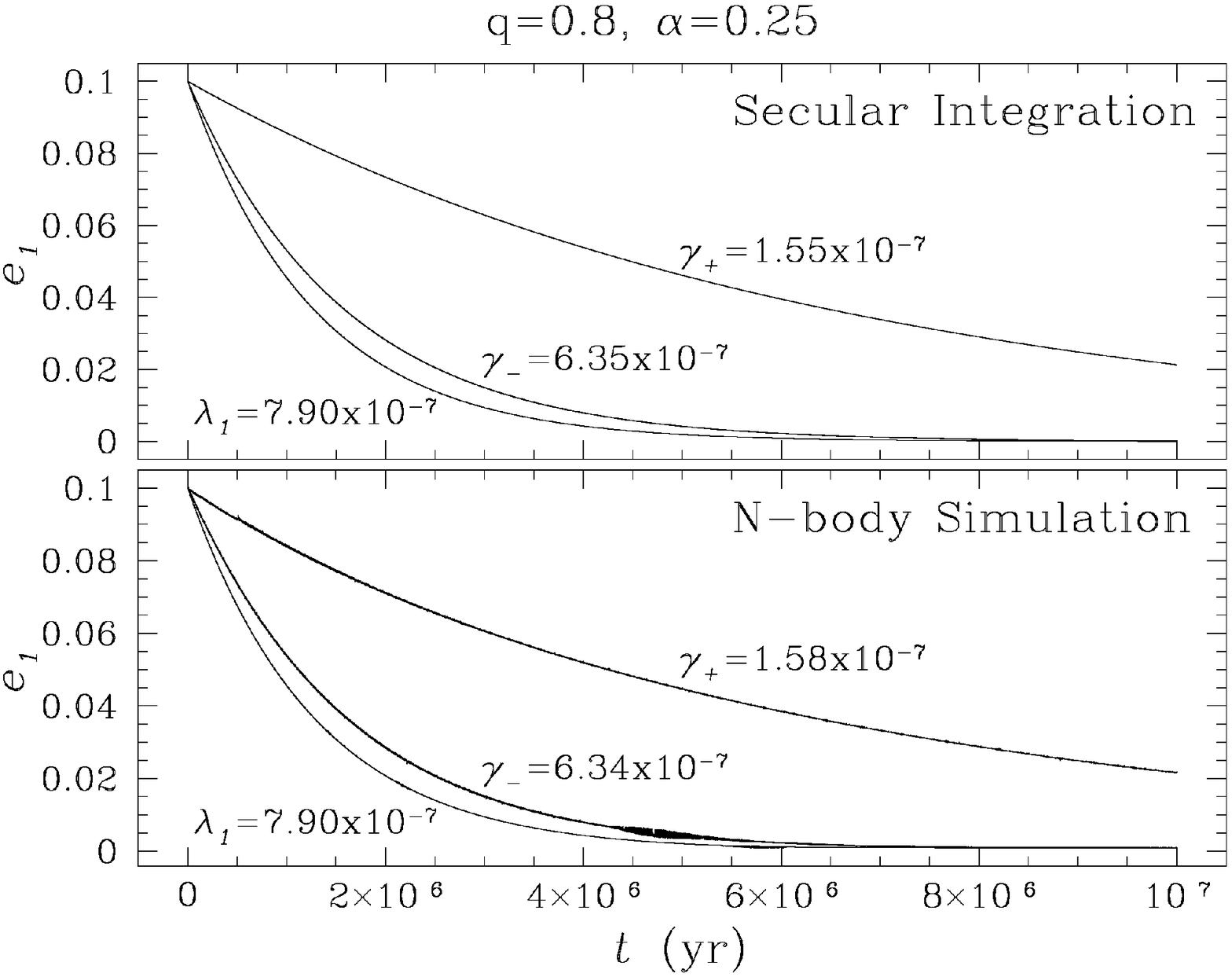} &
    \includegraphics[width=0.48\textwidth]{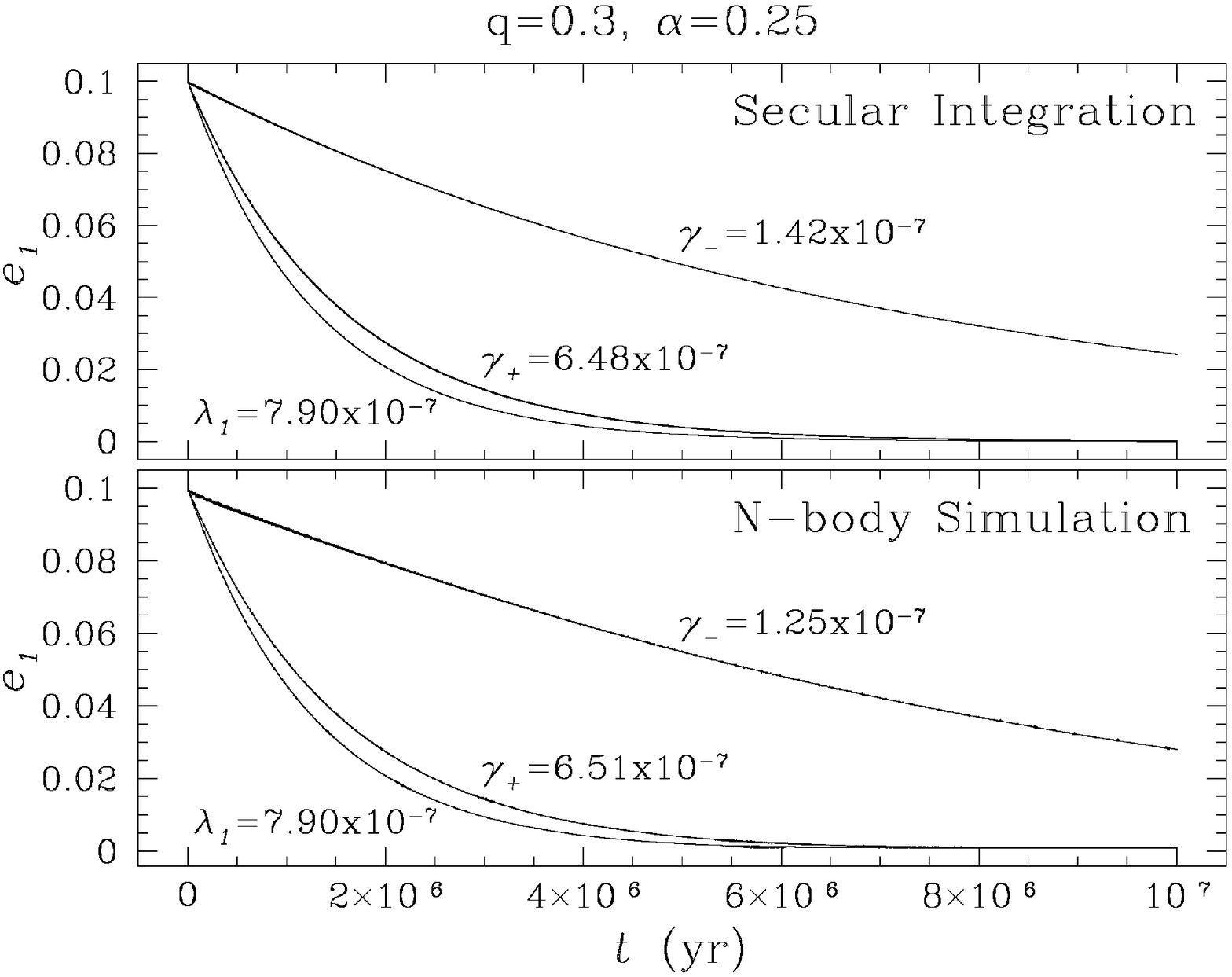} \\
    (a) & (b)\\
  \end{tabular}
%  \plottwo{plots/modedampqgsa.eps}{plots/modedampqlsa.eps}
  \caption[Eccentricity damping in secular modes]{Eccentricity damping
    of systems in different secular modes found by integration of the
    secular equations (top panels) and direct N-body simulations
    (bottom panels).  The plots show the eccentricity evolution of a 
    hot-Jupiter (1 Jupiter-mass planet at 0.05 AU from a 1
    solar-mass star) with a companion; $\gamma_+$ and $\gamma_-$
    represent the mode damping rates, which are e-folding times
    measured directly from the curves.  Also plotted is a single 
    hot-Jupiter subject to an artificial eccentricity damping with
    a rate $\lambda_1 = 7.90 \times 10^{-7}$ yr$^{-1}$.  (a) A 0.8
    Jupiter-mass companion is located at 0.2 AU ($q>\sqrt{\alpha}$),
    with predicted mode damping rates $\gamma_+ = 1.5499 \times
    10^{-7}$ yr$^{-1}$, $\gamma_- = 6.3501 \times 10^{-7}$ yr$^{-1}$
    (Equation~(\ref{eq.gamma})). (b) A 0.3 Jupiter-mass companion is located
    at 0.2 AU ($q<\sqrt{\alpha}$), with predicted $\gamma_+ = 6.4779
    \times 10^{-7}$ yr$^{-1}$ and $\gamma_- = 1.4221 \times 10^{-7}$
    yr$^{-1}$.}
  \label{fig.modedamp}
\end{figure*}

In Figure~\ref{fig.modedamp}, we compare our analytical results with
numerical integration of both the secular equations and the direct
N-body equations, with an artificial eccentricity damping added only
to the inner planet in all cases. The eccentricity evolution curves of
the inner planet are plotted for two different cases, and e-folding
rates are measured and labeled for all curves.
Figure~\ref{fig.modedamp}a illustrates a system with $q>\sqrt{\alpha}$.
The top panel shows results from secular equations, and the bottom
panel shows the corresponding N-body simulations.  Each panel plots
three curves: (1) the single planet case, in which the eccentricity of
the inner planet damps at rate $\lambda_1$, (2) a two-planet aligned
mode with damping rate $\gamma_+$, and (3) a two-planet anti-aligned
mode (damping rate $\gamma_-$).  For $q>\sqrt{\alpha}$,
Figure~\ref{fig.modedamp}a, eccentricities damp much faster in the
anti-aligned mode than in the aligned mode, as predicted by
Equation~(\ref{eq.gamma}).  A comparison between the top and bottom panel
shows close agreement (within 2\%) between full-scale N-body
simulation and integration of the approximate secular equations.
Damping rates predicted by Equation~(\ref{eq.gamma}) match the observed
secular decay rates almost perfectly.  Figure~\ref{fig.modedamp}b shows
a system with $q<\sqrt\alpha$, for which the aligned mode damps faster
than the anti-aligned mode.  The faster damping rate of the aligned
mode in the N-body simulation is within 0.5\% of the prediction, but
that of the slow anti-aligned mode, however, is $\sim 15\%$ off. This
discrepancy might be the result of unmodeled tidal perturbations to the inner
body's semi-major axis. The smaller $m_2$ of Figure~\ref{fig.modedamp}b
weakens the secular interaction, thereby emphasizing these tidal
effects.

The different damping rates for the two modes are particularly
interesting, especially for well-separated nearly-decoupled orbits for
which $\alpha$ and $\beta$ are small.  In this case, the
$4q\sqrt{\alpha}\beta^2$ term under the square root of
Equation~(\ref{eq.gamma}) is much smaller than the other term.  As a
result, if the eccentricity damping on one orbit is much faster than
on the other ($\lambda_1 \gg \lambda_2$), as in the case of tides, one
mode damps rapidly at nearly the single-planet tidal rate
$\lambda_1$. The system evolves quickly into a single mode which
decays substantially more slowly.

\begin{figure*}%[p]
  \noindent\includegraphics[width=\textwidth]{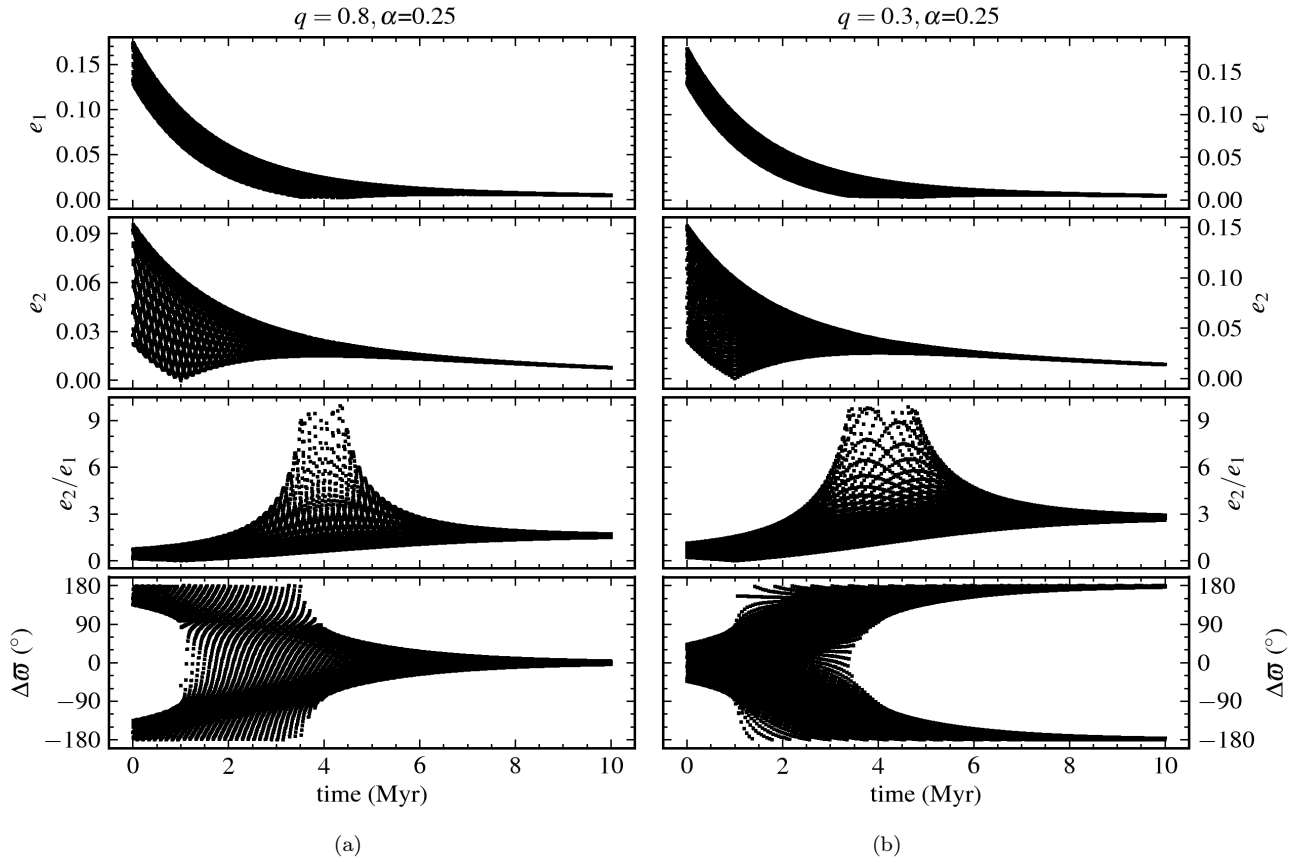}
  \hspace*{0.27\textwidth}(a) \hspace{0.37\textwidth}(b)
%  \plotone{plots/damp.eps}
  \caption{Secular evolution of the same systems shown in
    Figure~\ref{fig.modedamp}, but with different initial conditions so
    that the systems begin in mixed states.  (a) For
    $q>\sqrt{\alpha}$, the anti-aligned mode damps quickly at rate
    $\gamma_-$ from Figure~\ref{fig.modedamp}a, and the system evolves
    to the aligned mode ($\Delta\varpi\approx 0^\circ$). (b) For
    $q<\sqrt{\alpha}$, the aligned mode damps more rapidly and the
    system evolves to the anti-aligned mode ($\Delta\varpi\approx
    180^\circ$).}
  \label{fig.damp}
\end{figure*}

Because of the different damping rates for the two modes, a system will
evolve into a single mode even if it starts in a mixed state.
Figure~\ref{fig.damp} shows the eccentricity and apsidal angle evolution
of the systems depicted in Figure~\ref{fig.modedamp}, but with initial
conditions that lead to mixed states.  Figure~\ref{fig.damp}a shows the
case of $q>\sqrt{\alpha}$. Before 4~Myr, the system is in a mixed
state, so both eccentricities, as well as their ratio, oscillate (cf.
Figure~\ref{fig.soln}).  As the short-lived
anti-aligned mode damps away, the orbits begin to librate around
$\Delta\varpi\approx0^{\circ}$, the two eccentricities oscillate less and
less, and in the end, the eccentricity ratio settles to the aligned
mode ratio $|\eta_+|$ predicted by Equation~(\ref{eq.eta}).
Figure~\ref{fig.damp}b shows the corresponding plots for
$q<\sqrt{\alpha}$.  The orbital elements undergo similar evolution,
except that the aligned mode damps quickly and the system ends up in
the anti-aligned mode.

\subsection{Apsidal Circulation and Libration} \label{sec.apsidal}

\begin{figure*}%[p]
  \includegraphics[width=\textwidth]{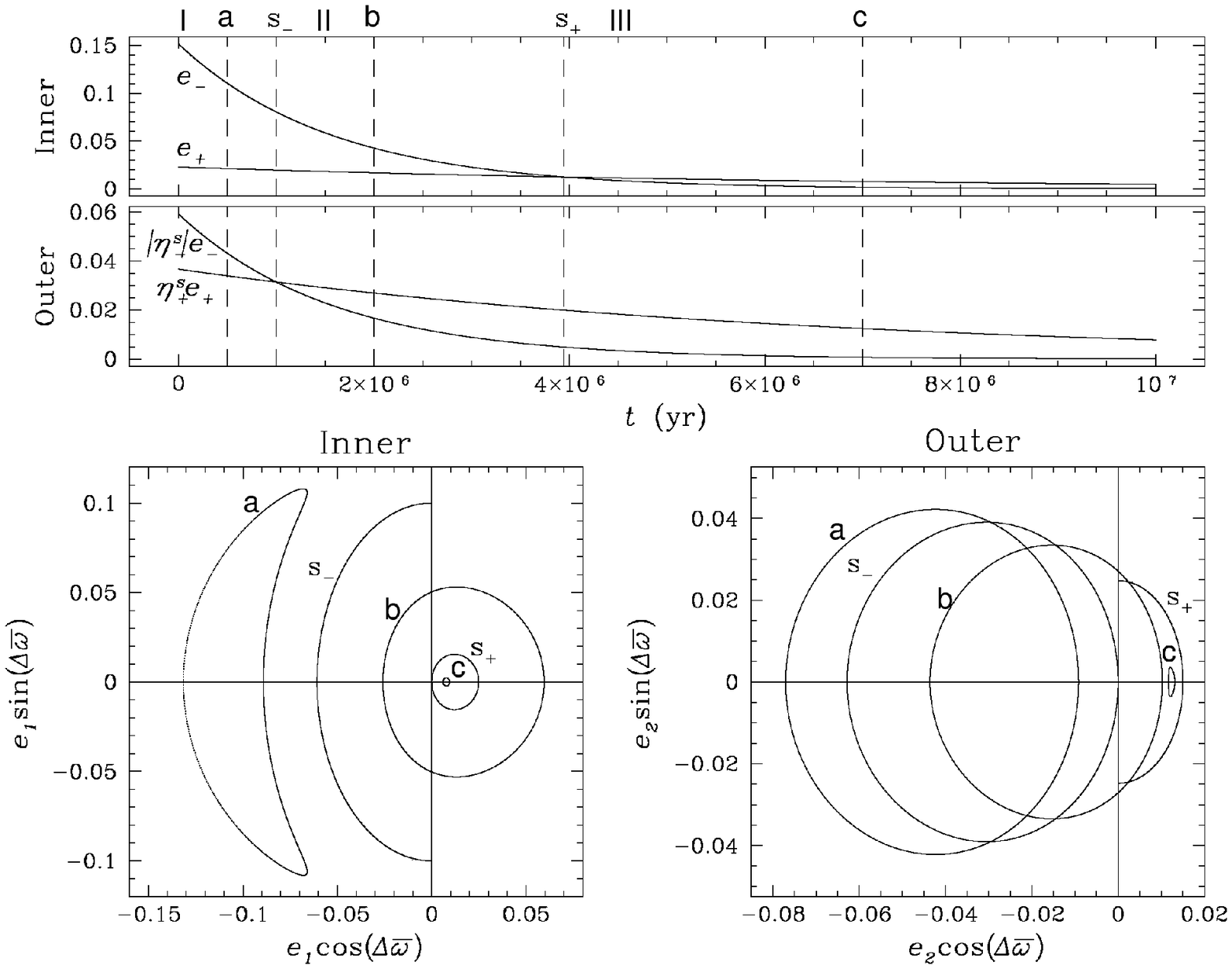}
%  \plotone{plots/ssphasep.eps}
  \caption[Apsidal state evolution for $m_1^2a_1 < m_2^2a_2$
    case]{Evolution of the apsidal states during eccentricity damping.
    The top two panels show the time evolution of the eccentricity
    components for the system in Figure~\ref{fig.damp}(a) for which
    $m_1^2a_1 < m_2^2a_2$ and the anti-aligned mode damps fastest.
    The inner orbit has an aligned component $e_+$ and an anti-aligned
    one $e_-$, while $\eta_+^se_+$ and $|\eta_-^s|e_-$ are the
    components for the outer orbit.  Two circulation-libration
    separatrices ($s_-$ and $s_+$) divide the evolution curves into
    three parts: anti-aligned libration (region I), circulation
    (region II), and aligned libration (region III).  The bottom
    panels show phase diagrams of the inner and outer orbits on the
    complex $e\exp(i\Delta\varpi)$ plane at the corresponding points
    indicated in the top panels.  The shape of the phase curves
    depends on the relative strength of the two components for each
    orbit.  Banana shapes result from the large difference between the
    two components: $e_- \ll e_+$ at (a) and $|\eta_-^s|e_- \ll
    \eta_+^se_+$ at (c).}
  \label{fig.ssphasep}
\end{figure*}

In Figure~\ref{fig.damp}, the apsidal motion of the two orbits changes from
libration of $\Delta\varpi$ about $0^{\circ}$ or $180^{\circ}$
to circulation of $\Delta\varpi$ through a full $360^{\circ}$, and
to libration again during the eccentricity damping. In order to
understand what determines the apsidal state of the orbits, we plot
the aligned and anti-aligned components of the eccentricities for
Figure~\ref{fig.damp}a in the top two panels of Figure~\ref{fig.ssphasep}.
Recall that the total eccentricities of the orbits at any time can be
obtained from the components as illustrated in Figure~\ref{fig.soln}.
The lower panels in the figure show the phase diagrams of both orbits
($e\cos(\Delta\varpi)$ versus $e\sin(\Delta\varpi)$) at five different
time indicated in the top two panels.  The two orbits move
along the phase curves, which themselves change slowly over time.  Two
critical instants, labeled $s_-$ and $s_+$, are circulation-libration
separatrices, which represent the transitions of the apsidal state
from anti-aligned libration to circulation ($s_-$), and from
circulation to aligned libration ($s_+$).  These two points divide the
evolution curves into three regions.

In region I ($t<10^6$ yr), the anti-aligned components are stronger
than the aligned ones for both orbits ($e_->e_+$ and $|\eta_-^s|e_- >
\eta_+^se_+$).  In the phase plots, both the $e_1$ and the $e_2$
curves are closed and stay on the left side of the vertical
$e\sin(\Delta\varpi)$ axis, indicating the libration of $\Delta\varpi$
about $180^{\circ}$.  With the decrease in the amplitudes of all
components, especially the faster damping of the anti-aligned ones,
the curves move closer toward the origin, resulting in an increased
libration width of $\Delta\varpi$.

The anti-aligned separatrix $s_-$ crossing occurs at $t=10^6$ yr,
when the two components for the outer orbit are equal ($|\eta_-^s|e_-
= \eta_+^se_+$) and $e_2$ might drop to zero, resulting in a phase curve
for the orbit that is tangent to the vertical axis at
the origin ($s_-$ curve on the inner phase plot of
Figure~\ref{fig.ssphasep}).  The phase curve for $e_1$ at $s_-$ is a
half-oval whose straight edge includes the origin.  Accordingly, when
$e_2$ drops to zero, $\Delta\varpi$ jumps from $90^{\circ}$ to
$-90^{\circ}$ for the largest possible full libration amplitude of
$180^{\circ}$.

As the system moves past $s_-$, the phase curves for both orbits
enclose the origin and circulation results.  The circulation region
(II) is located between the two separatrices (i.e., when $10^6$ yr $<
t < 3.95\times10^6$ yr), where the anti-aligned component of the
inner orbit is stronger than the aligned one ($e_->e_+$), while it is
weaker for the outer orbit ($|\eta_-^s|e_- < \eta_+^se_+$).  With the
continuous fast damping of the anti-aligned mode, the system crosses
the aligned separatrix $s_+$ at $t=3.95\times10^6$ yr, when the two
components for the inner orbit are equal ($e_-=e_+$).  The two
separatrices occur at those times when each phase curve in
Figure~\ref{fig.ssphasep} touches the origin.

After $s_+$, both phase curves are to the right of the
vertical axis, indicating libration about the aligned
mode (region III).  The two anti-aligned components are both
significantly damped and the system now has both $e_-<e_+$ and
$|\eta_-^s|e_- < \eta_+^se_+$. 

\begin{figure}%[p]
  \plotone{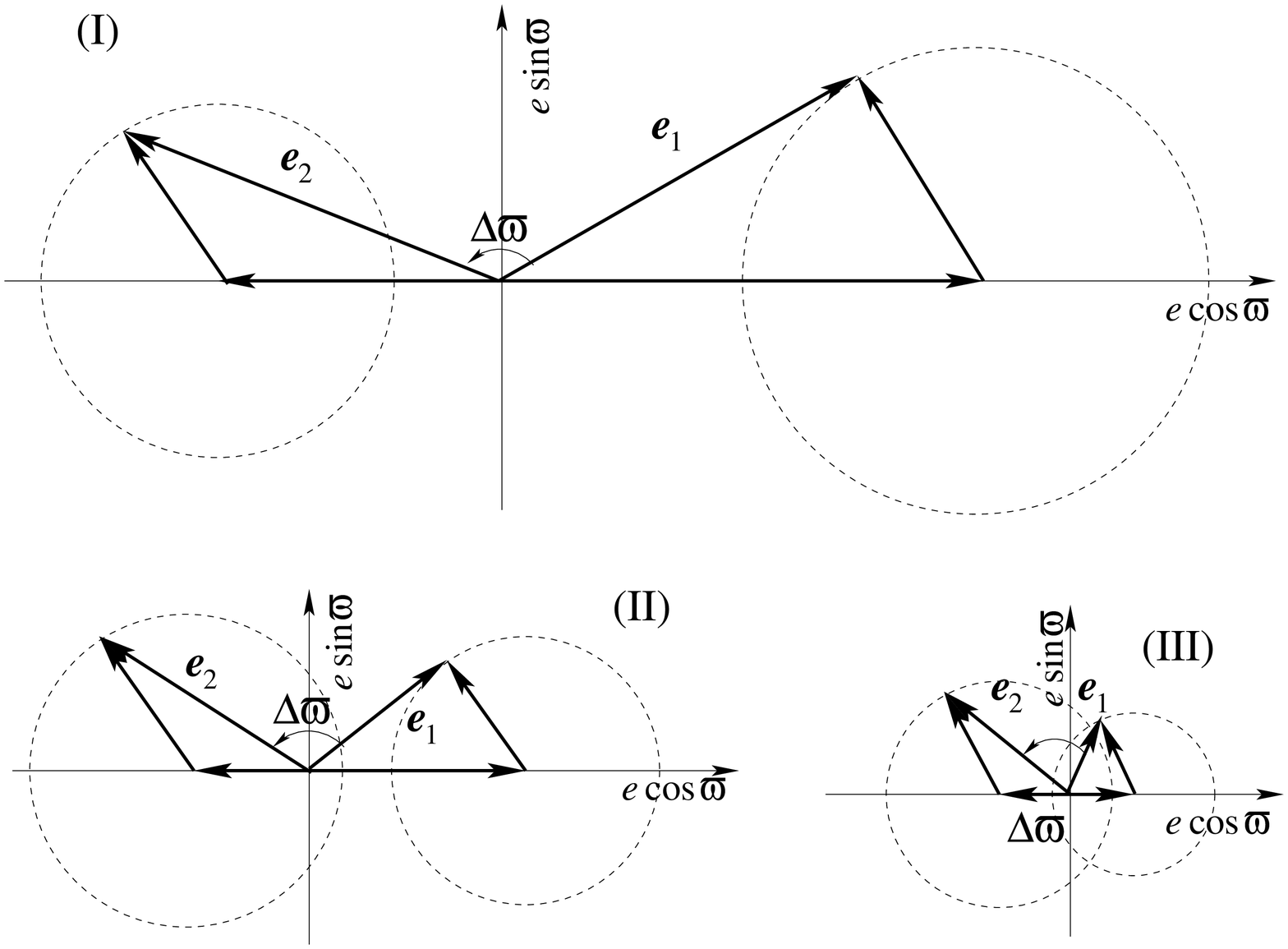}
  \caption[Eccentricity component diagrams for different apsidal
  states]{Eccentricity component diagrams for different regions in
    Figure~\ref{fig.ssphasep}.  These diagrams are similar to
    Figure~\ref{fig.soln}, but now shown in a frame rotating at the same
    rate as the anti-aligned mode so that the horizontal vectors are
    $\mathrm{{\bf e_-}}$ and $\mathrm{\eta_-{\bf e_-}}$.  Here, the anti-aligned
    mode damps faster than the aligned mode ($\gamma_->\gamma_+$) so
    that the circles move horizontally toward the origin faster than
    their radii shrink.  The system starts with $\Delta\varpi \approx
    0^\circ$ in region I and evolves to $\Delta\varpi\approx
    180^\circ$ in region III.}
  \label{fig.lib}
\end{figure}

The geometry of the orbits can also be illustrated with a component
diagram similar to Figure~\ref{fig.soln}, but in a frame rotating at the
same rate as the anti-aligned mode (Figure~\ref{fig.lib}).  
In this rotating frame, the aligned component vectors always rotate
clockwise because their precessions are slower than those of the
anti-aligned ones.  Evolution in these coordinates can be visualized
as circles whose radii and horizontal distances from the origin shrink
at the different rates $\gamma_+$ and $\gamma_-$.  Since the
anti-aligned components initially dominate the aligned ones (region
I), the $\mathbf{e}_1$ vector stays on the right side of the
vertical axis, and the $\mathbf{e}_2$ vector on the left side
in Figure~\ref{fig.lib}.
When the aligned components are parallel to the vertical axis, the
angle between $\mathbf{e}_1$ and $\mathbf{e}_2$, $|\Delta\varpi| >
90^{\circ}$, is at minimum; thus the orbits librate about
$\Delta\varpi=180^{\circ}$ (anti-aligned libration).  When the
$\mathbf{e}_2$ circle moves to enclose the origin and the
$\mathbf{e}_1$ circle is still confined in the first and fourth
quadrants, the system reaches circulation region II.  This geometry
enables $\Delta\varpi$ to cycle through a full $360^{\circ}$
(Figure~\ref{fig.lib}).  Last, when the anti-aligned components are
sufficiently damped and both circles contain the origin, the system
goes to aligned mode libration (region III).  Now the maximum value of
$\Delta\varpi <90^{\circ}$ occurs when the two aligned components are
parallel to the vertical axis, and the orbits librate about
$\Delta\varpi=0^{\circ}$.

\begin{figure*}%[p]
  \includegraphics[width=\textwidth]{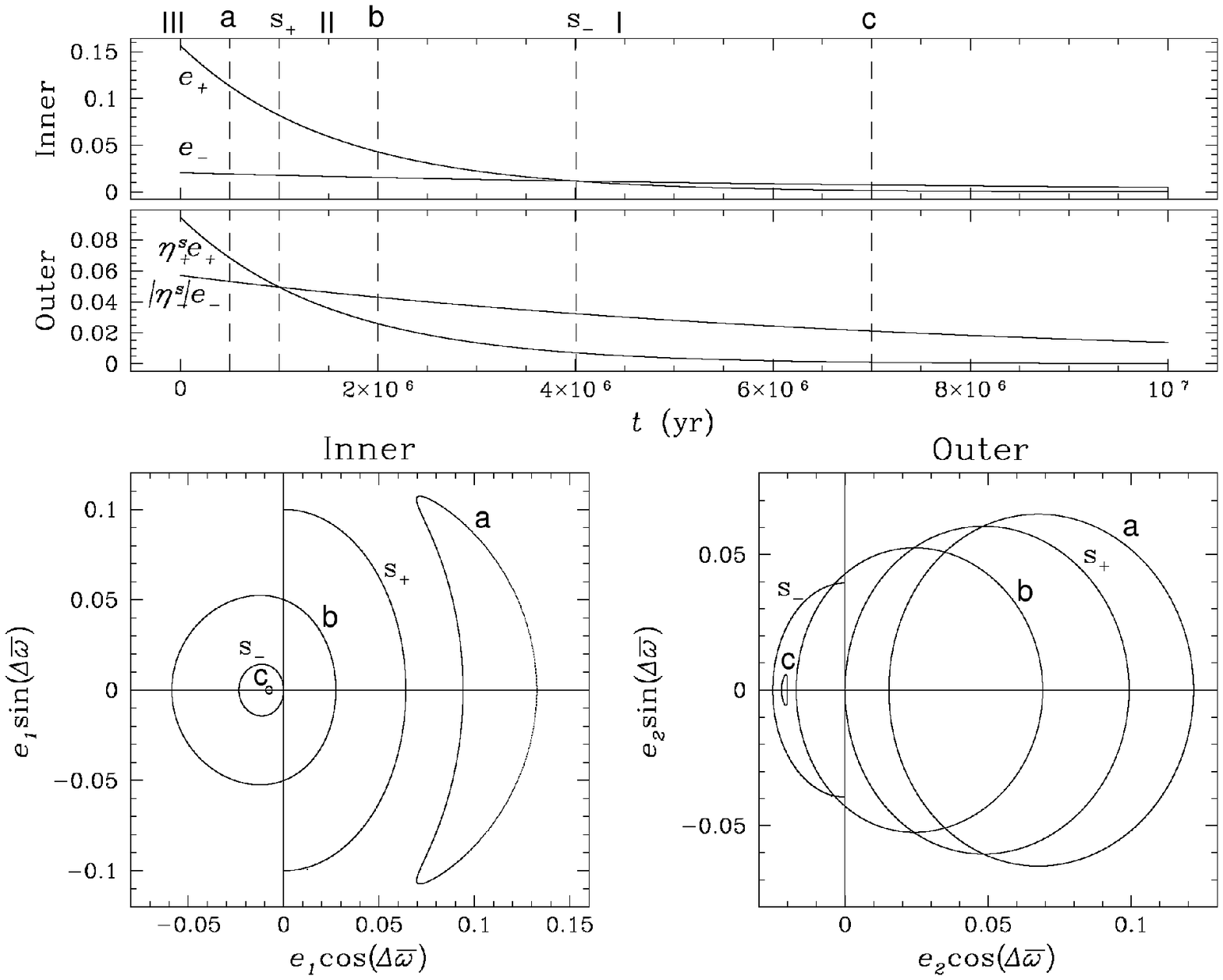}
%  \plotone{plots/ssphasem.eps}
  \caption[Apsidal state evolution for $m_1^2a_1 > m_2^2a_2$
    case]{Evolution of the apsidal state during eccentricity damping.
    Similar to Figure~\ref{fig.ssphasep}, but using data from
    Figure~\ref{fig.damp}(b) for which $m_1^2a_1 > m_2^2a_2$. The aligned
    mode damps fastest and the system moves from aligned libration to
    circulation, and finally to anti-aligned libration.}
  \label{fig.ssphasem}
\end{figure*}

Figure~\ref{fig.ssphasem} shows the case of Figure~\ref{fig.damp}b, where
the two aligned components are initially stronger and the system
starts in the aligned libration region III.  The two orbits evolve to
cross the aligned separatrix $s_+$ into the circulation region II, and
then pass the anti-aligned separatrix $s_-$ to reach the final
anti-aligned libration region I.  The equivalent of Figure~\ref{fig.lib}
for this system would show two circles that initially encompass the
origin; here the radii of the circles would shrink faster than the
distances of their centers from the origin.

In conclusion, the apsidal state of a two-planet secular system
depends on the sign of the simple product:
\[
P = (e_+-e_-)(\eta_+^se_+ + \eta_-^se_-).
\]
Libration occurs when the same mode components are stronger for both
orbits ($P > 0$), and circulation occurs when one mode is stronger for
the inner orbit, but weaker for the outer one ($P < 0$). This result
is in full agreement with a slightly more complicated formula given by
\citet{barnes2006478}.

Eccentricity damping is effective in changing the apsidal state of the
orbits because the two modes damp at different rates.
Eccentricity excitation is equally capable of moving the two orbits
across libration-circulation separatrices.  This can be easily
visualized by running the plots in Figures~\ref{fig.ssphasep} and
\ref{fig.ssphasem} backward in time.  For systems with
$m_1^2a_1<m_2^2a_2$ (see Figure~\ref{fig.ssphasep}), eccentricity
excitation would eventually bring the orbits into anti-aligned
libration (region I), while eccentricity damping brings them into
aligned libration (region III).  The opposite is true for systems with
$m_1^2a_1>m_2^2a_2$ (Figure~\ref{fig.ssphasem}).  All mechanisms that
change eccentricities slowly cause planetary systems to move toward
apsidal libration.

\subsection{Additional Apsidal Precessions} \label{sec.gr}

%Extrasolar ``hot-Jupiters'' are close enough to their host stars that
In Sections~\ref{sec.damping} and \ref{sec.apsidal}, we have considered
tides as an orbital circularization mechanism.  Tides, however, also
cause apsidal precession \citep[see, e.g.,][]{ragozzine20091778}:
\begin{eqnarray}
\dot{\varpi}^{_{T,p}} &=& \frac{15}{2}k_{2p}\left(\frac{R}{a}\right)^5 \frac{m_*}{m} n, \nonumber \\
\dot{\varpi}^{_{T,*}} &=& \frac{15}{2}k_{2*}\left(\frac{R_*}{a}\right)^5 \frac{m}{m_*} n. \nonumber
\end{eqnarray}
Here, $\dot{\varpi}^{_{T,p}}$ and $\dot{\varpi}^{_{T,*}}$ are the tidal
precession rates due to planetary and stellar tides, respectively; $k_{2p}$
is the Love number of the planet; and $k_{2*}$ is that of the
star. These expressions assume that tidal bulges are directly
underneath the distant body (non-dissipative tides) but could adjusted
to account for slight angular offsets in the tidal bulges (dissipative
tides). 

For close-in exoplanets, orbital precession caused by GR effects is also important.  
To lowest order in eccentricity, the precession rate \citep[e.g.,][]{danby1988fcm} 
is the following:
\[
\dot{\varpi}^{_{GR}}=\frac{3 \, a^2 n^3}{c^2},
\]
where $c$ is the speed of light.  

In addition, the rotational bulges raised on the planet and its star also
lead to orbital precession, with respective rates
\citep{ragozzine20091778}:
\begin{eqnarray}
\dot{\varpi}^{_{R,p}} &=& \frac{k_{2p}}{2} \left(\frac{R}{a}\right)^5 \frac{m_*}{m} \frac{\Omega^2}{n^2} n, \nonumber \\
\dot{\varpi}^{_{R,*}} &=& \frac{k_{2*}}{2} \left(\frac{R_*}{a}\right)^5 \frac{\Omega^2_*}{n^2} n. \nonumber
\end{eqnarray}
Here, $\Omega$ and $\Omega_*$ are the spin rates of the planet and the
star, respectively.

To account for these additional orbital precessions, we define a
dimensionless quantity:
\[
\kappa=\frac{\dot{\varpi}^{_{T,p}}_1 + \dot{\varpi}^{_{GR}}_1 + \dot{\varpi}^{_{R,p}}_1 +
\dot{\varpi}^{_{T,*}}_1 +  \dot{\varpi}^{_{R,*}}_1}{\sigma},
\]
which is the ratio of the sum of all additional precessions to the
characteristic secular precession. These additional precessions add
extra terms to Equation~(\ref{eq.lang}), which now reads as follows:
\[
\dot{\varpi}_j = + \frac{1}{n_j a_j^2 e_j}
  \frac{\partial\mathcal{R}_j}{\partial e_j} 
  + \kappa \sigma.
\]
Since $\kappa$ is independent of $e$ for small eccentricities, the
extra terms do not change the form of Equation~(\ref{eq.hdot}), and so all
discussion of the general secular modes still holds.  In particular,
the system still has aligned and anti-aligned modes and the two modes
damp separately.  The mode frequencies, damping rates, and
eccentricity ratios, however, need to be revised.  Now the diagonal
terms of the coefficient matrix $A_{jk}$ should be adjusted to the following:
\[
A= \sigma \left\{ \begin{array}{cc}
     q + \kappa +i \xi_1 & - q \beta \\
     - \sqrt{\alpha} \beta & \sqrt{\alpha}(1+\alpha^2\kappa) + i \xi_2
     \end{array} \right\},
\]
which gives the new mode frequencies and eccentricity ratios:
\footnotesize
\begin{eqnarray}
  g_{\pm}^s &=& \frac{1}{2} \, \sigma \left\{ (q+ \kappa)
    +\sqrt{\alpha}(1+\alpha^2\kappa) \mp \sqrt{[q +\kappa
    -\sqrt{\alpha}(1+\alpha^2\kappa)]^2 + 4\,q\sqrt{\alpha} \beta^2}
    \right\}, \label{eq.gsgr} \\
  \eta_{\pm}^s &=& \frac{q +\kappa -\sqrt{\alpha}(1+\alpha^2\kappa)
    \pm \sqrt{[q +\kappa -\sqrt{\alpha}(1+\alpha^2\kappa)]^2 +
    4\,q\sqrt{\alpha} \beta^2}} {2\,q\beta}, \label{eq.etasgr}\\
  g_{\pm} &=& g_{\pm}^s \pm \frac{q \sqrt{\alpha} \beta^2}
    {\{[q+\kappa -\sqrt{\alpha}(1+\alpha^2\kappa)]^2 + 4 q
    \sqrt{\alpha} \beta^2\}^{3/2}} \, \sigma \, (\xi_1-\xi_2)^2, \label{eq.ggr}\\
  \gamma_{\pm} &=& \frac{1}{2} \left[\lambda_1+\lambda_2 \pm \frac{
    \sqrt{\alpha}(1+\alpha^2\kappa) - (q+\kappa)} {\sqrt{[q+\kappa
    -\sqrt{\alpha}(1+\alpha^2\kappa)]^2 + 4 q \sqrt{\alpha}
    \beta^2}}(\lambda_1-\lambda_2)\right] , \label{eq.gammagr}\\ 
  \eta_{\pm} &=& \eta_{\pm}^s \left\{ 1 \pm i \frac{\xi_1-\xi_2}
    {\sqrt{[q+\kappa - \sqrt{\alpha}(1+\alpha^2\kappa)]^2 + 4\, q
    \sqrt{\alpha} \beta^2} } \right\}. \label{eq.etagr}
\end{eqnarray}
\normalsize
Note that these equations can be obtained from Equations~(\ref{eq.gs}), 
(\ref{eq.etas}), (\ref{eq.g}), (\ref{eq.gamma}), and (\ref{eq.eta}) with the
simple substitution $ (q\pm\sqrt{\alpha}) \rightarrow [q+\kappa \pm
\sqrt{\alpha}(1+\alpha^2\kappa)]$. 

These additional apsidal precessions increase both secular rates
$g_{\pm}^s$ (Equation~\ref{eq.gsgr}) since they cause the orbits to precess
in the same direction as the secular interaction does.  The mode
eccentricity ratio $|\eta^s_+|$ (Equation~\ref{eq.etasgr}) increases
significantly with any source of precession that favors the inner
orbit, indicating that it is more difficult to force the eccentricity
of a rapidly precessing inner orbit in the aligned mode. The ratio
$|\eta^s_-|$, however, decreases slightly; increasing the precession
of the inner orbit in the anti-aligned mode actually strengthens
secular coupling. As for the mode damping rates
(Equation~\ref{eq.gammagr}), additional precession decreases the aligned
mode damping rate, but increases that of the anti-aligned mode.  It
also decreases $\epsilon$, the deviation angle of the mode apsidal
lines from perfect alignment (Equation~\ref{eq.etagr}).

%================================================================================================
% SECTION 3
%\section{Results} \label{sec.results}
\section{Applications to the Observed Systems} \label{sec.results}

%In this section, 

% We now use the theory developed in Section~\ref{sec.model} to
% analyze the current status of a two-planet system (HAT-P-13) in
% Section~\ref{sec.2plsystems}.  We then discuss how our equations can
% be used to predict possible companions of the observed single close-in
% planet systems with (Section~\ref{sec.1ptrend}) and without linear
% trends (Section~\ref{sec.1pnotrend}).

We now apply the aforementioned theory to close-in exoplanets,
beginning with known two-planet systems.  We then proceed to systems
in which there is linear trend in the star's radial velocity (RV) that might 
signal the presence of an outer companion and finally we consider
systems with no hint of a companion.

%\begin{sidewaystable}%\footnotesize
%\begin{landscape}
\begin{table*} 
  \caption{Properties of Planets and Stars Discussed in This Paper}
  \label{tab_orb_el}
  \hspace{-0.15in}
%  \hspace{-1in}
%  \vspace{0.1in}
  \begin{tabular}{lcccccccc}
 \hline \hline
    Planet & $m_p$ ($m_J$) & $R_p$ ($R_J$) & $a$ (AU) & $e$ & $\omega$ ($^{\circ}$) & $m_*$ ($m_{\sun}$) & Age (Gyr) & $\tau_e\,$(Gyr)\\ \hline
% 2pl systems
HAT-P-13\,b & $0.85\pm0.0354$ & $1.28\pm0.079$ & $0.0427\pm0.000875$ & $0.013\pm0.0041$ & $210^{+27}_{-36}$ & $1.2^{+0.05}_{-0.1}$ & $5^{+2.5}_{-0.8}$ & $\sim0.125$\\
HAT-P-13\,c & $14.3\pm0.691$ &  & $1.23\pm0.0251$ & $0.662\pm0.0054$ & $175.3\pm0.35$ & $1.2^{+0.05}_{-0.1}$ & $5^{+2.5}_{-0.8}$ & \\
HD\,187123\,b & $0.51\pm0.0173$ & $1^*$ & $0.0421\pm0.000702$ & $0.010\pm0.00593$ & $24.5$ & $1.04^{+0.026}_{-0.024}$ & $5.33$ & $\sim0.3$\\             
HD\,187123\,c & $1.9\pm0.152$ &  & $4.8\pm0.367$ & $0.25\pm0.0334$ & $240\pm18.6$ & $1.04^{+0.026}_{-0.024}$ & $5.33$ &
\\ \hline
% 1pl systems + LT
GJ\,436\,b & $0.073\pm0.00318$ & $0.3767^{+0.0082}_{-0.0092}$ & $0.0287\pm0.000479$ & $0.16\pm 0.019$ & $351\pm1.2$ & $0.45^{+0.014}_{-0.012}$ & $6^{+4}_{-5}$ & $\sim2$\\
BD\,-10\,3166\,b & $0.43\pm0.0174$ & $1^*$ & $0.0438\pm0.000730$ & $0.02^{+0.042}_{-0}$ & $334$ & $0.92^{+0.046}_{-0.024}$ & $4.18$ & $\sim0.42$\\
HAT-P-26\,b & $0.059\pm0.00718$ & $0.565\pm0.052$ & $0.0479\pm0.000798$ & $0.12\pm 0.06$ & $100\pm165$ & $0.82\pm0.033$ & $9^{+3}_{-4.9}$ & $\sim2.6$\\
WASP-34\,b & $0.58\pm0.0285$ & $1.22^{+0.11}_{-0.08}$ & $0.052\pm0.00120$ & $0.04\pm 0.0012$ & $320\pm20.9$ & $1.01\pm0.07$ & $6.7^{+6.9}_{-4.5}$ & $\sim0.7$\\
HD\,149143\,b & $1.33\pm0.0784$ & $1^*$ & $0.053\pm0.00147$ & $0.016\pm 0.01$ & $0$ & $\sim1.2\pm0.1$ & $7.6\pm1.2$ & $\sim2.75$\\ 
\hline
% 1pl systems
HAT-P-21\,b & $4.1\pm0.173$ & $1.024\pm0.092$ & $0.0495\pm0.000825$ & $0.23\pm 0.016$ & $309\pm3$ & $0.95\pm0.042$ & $10.2\pm2.5$ & $\sim6$\\   
HAT-P-23\,b & $2.1\pm0.122$ & $1.368\pm0.09$ & $0.0232\pm0.000387$ & $0.11\pm 0.044$ & $120\pm25$ & $1.13\pm0.035$ & $4\pm1$ & $\sim0.005$\\
HAT-P-32\,b & $1.0\pm0.169$ & $2.037\pm0.099$ & $0.0344\pm0.000574$ & $0.16\pm 0.061$ & $50\pm29$ & $1.18^{+0.043}_{-0.07}$ & $3.8^{+1.5}_{-0.5}$ & $\sim0.005$\\
HAT-P-33\,b & $0.8\pm0.117$ & $1.83\pm0.29$ & $0.050\pm0.00115$ & $0.15\pm 0.081$ & $100\pm119$ & $1.40\pm0.096$ & $2.4\pm0.4$ & $\sim0.06$\\
HD\,88133\,b & $0.30\pm0.0270$ & $1^*$ & $0.0472\pm0.000786$ & $0.13\pm 0.072$ & $349$ & $1.2\pm0$ & $9.56$ & $\sim0.38$\\
%
%    HIP 14810b & $6.674\pm0.002$ & $3.9\pm0.6$ &
%    $0.069\pm0.004$ & $0.147\pm 0.006$ & $159\pm2$ & $\sim0.99$ & $\sim 4$ \\
%    HIP 14810c & $95.285\pm0.002$ & $0.8\pm0.1$ &
%    $0.41\pm0.02$ & $0.409\pm0.006$ & $354\pm2$ & $\sim0.99$ & $\sim 4$ \\
%    GJ 436b & $2.64385 \pm 0.00009$ & $0.0713\pm0.006$ &
%    $0.029\pm0.002$ & $0.16\pm0.02$ & $351\pm1$ & $0.44\pm0.04$ &
%    $>3$ \\
%    GJ 674b & $4.6938\pm0.007$ & 0.037 & 0.039 & $0.20\pm0.02$&
%    $143\pm6$ & 0.35 & 0.1-1 \\
    \hline \hline
  \end{tabular}
  \tablerefs{Data for HAT-P-13 system are from \citet{winn2010575};
    HD\,187123 from \citet{wright20091084}; HAT-P-21 and HAT-P-23 from
    \citet{bakos2011116}; HAT-P-26 \citet{hartman2011138}; HAT-P-32 and HAT-P-33 from
    \citet{hartman201159}; HD\,88133 and BD\,-10\,3166 from
    \citet{butler2006505}; GJ 436 from \citet{maness200790}; WASP-34
    from \citet{smalley2011130}; HD\,149143 from \citet{fischer20061094}. 
    The planetary mass $m_p$ assumes the lower limit $m_p\sin i$ for listed planets 
    without measured planetary radius $R_p$.   
    The circularization timescale $\tau_e$ is estimated by integrating a set of 
    tidal equations \citep[e.g.,][]{matsumura20101995} 
    from the tabulated eccentricity to $e=10^{-4}$. An asterisk signifies an assumed value.}
%GJ 674 from \citet{bonfils2007293}.}
\end{table*}
%\end{landscape}
%\end{sidewaystable}
%
%
%
%\subsection{A Test of the Theory: the HIP 14810 System}
%\subsection{Test of the Theory: Two-planet Systems} \label{sec.2plsystems}
\subsection{Two-planet Systems}  \label{sec.2plsystems}

There are more than 30 multi-planet systems that host one or more close-in
planets with the orbital period $P_{\rm orb}\lesssim 20$ days. 
%have been discovered.  
Out of these, many systems including Gliese 876, 55 Cnc, and
$\upsilon$ And have three or more planets, which makes apsidal
analysis more complicated \citep[e.g.,][]{barnes200653}.  Moreover,
most Kepler-detected planets have unknown eccentricities, and are not
the best candidates for our analysis.  After removing multiple planet systems and those with poorly determined eccentricities, we are
left with eight two-planet systems, with which we can test the linear tidal model.

First, we compute eccentricity damping timescales for the inner
planets ($\tau_e$), and compare $\tau_e$ with the stellar age
($\tau_{\rm Age}$) to determine whether an orbit has had sufficient
time to circularize. For all systems, we adopt conventional planetary
and stellar tidal quality factors $Q_p=10^5$ and $Q_*=10^6$,
respectively \citep[see, e.g.,][and references
therein]{jackson20081396,matsumura20101995}.  We convert these 
to modified tidal quality factors as $Q^{\prime}=1.5Q/k_2$. 
For giant planets in our solar system
(Jupiter, Saturn, Uranus, and Neptune), the measured gravitational
moments agree well with an $n\sim1$ polytrope
\citep{hubbard197442,bobrov1978489} which corresponds to the
Love number $k_2=0.52$ \citep{motz1952562}.
%the apsidal motion constant of $k=0.26$ \citep[][]{motz1952562}, or the Love number of $k_2=2k=0.52$.  
For most stars, on the other hand, the $n=3$ polytrope is a good
approximation \citep[e.g.,][]{horedt2004306}, which yields
$k_2=0.028$ \citep{motz1952562}.
%$k=0.014$ \citep{motz1952562} and thus $k_2=0.028$.  
We finally obtain $Q^{\prime}_p=2.88\times10^5$ and
$Q^{\prime}_*=5.36\times10^7$.  We estimate the eccentricity damping
timescale $\tau_e$ by integrating a set of tidal equations based on
the equilibrium tide model from the measured planetary eccentricity
down to $e=10^{-4}$, and further assume that the tidal quality factors
evolve proportional to the inverse of the mean motion $Q\propto 1/n_1$
\citep[see, e.g.,][]{matsumura20101995}.  Some of our results appear in
the final column of Table~\ref{tab_orb_el}. Note that these simulation
results compare favorably to the simpler approximation of
Equation~(\ref{eq.erate}) when one properly accounts for the multiple
e-folding times needed to damp eccentricities to $e=10^{-4}$.
%We Note that, although we do not
%take account of the orbital migration in our secular model, the
%eccentricity damping time is estimated in the model that includes the
%orbital decay due to tidal effects.  
Only 2 of the 8 two-planet systems (HD\,187123 and HAT-P-13) have 
short tidal circularization times compared with the stellar ages; physical
and orbital parameters of these two systems can be found at the top of
Table~\ref{tab_orb_el}.  Since $\tau_e << \tau_{\rm Age}$, the systems
are likely to have been significantly modified by tides.

Next, we check the strength of secular interactions for the two
systems. In the HD~187123 system, the inner planet is at $\sim
0.04\,$AU while the outer one is at $\sim 5\,$AU.  With this
configuration, the inner planet is strongly bound to the star and its
interaction with the outer planet is weak.  The inner and outer
planets of HAT-P-13, however, are more closely spaced at $\sim0.04$
and $\sim1.2\,$AU, respectively. Furthermore, since $m_2 >> m_1$ and
$e_2=0.66$ is large (Table~\ref{tab_orb_el}), the secular forcing of
the inner planet is substantial. Thus, out of the eight systems, only
HAT-P-13 has both strong secular interactions and a short tidal
damping time; it is, accordingly, the best test case for our theory.

%Out of these two systems, HD\,187123 has weak secular interactions,
%since the inner planet is at $\sim 0.04\,$AU while the outer one is at
%$\sim 5\,$AU.  On the other hand, HAT-P-13 has strong enough secular
%interactions between planets to perturb each other's orbits.  Since
%the tidal quality factors could be very different from our
%assumptions, we have also checked whether the other six two-planet
%systems have strong secular interactions.  In all of these systems
%(BD-082823, HD\,190360, HD\,215497, HD\,217107, HD\,38529, and
%HD\,47186), the secular interactions turn out to be weak.  For the
%rest of this subsection, we will concentrate on the HAT-P-13 system.

\begin{figure}%[p]
  \plotone{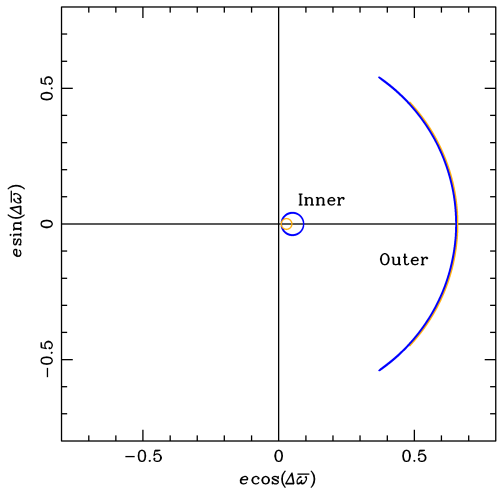}
  \vspace{-0.5in}
  \caption{Current apsidal state of the HAT-P-13 system.  We show
    the orbits of both planets on the $e\exp(i\Delta\varpi)$ plane,
    similar to the lower plots in Figures~\ref{fig.ssphasep} and
    \ref{fig.ssphasem}.  The orange curves represent the solution of
    secular equations, and the blue curves are obtained from an N-body
    simulation.  The orbits librate about $\Delta\varpi= 0^{\circ}$,
    with a libration amplitude $\sim 43^{\circ}$ predicted by secular
    theory and $\sim 57^{\circ}$ measured from the N-body
    simulation. The narrowness of the outer planet's arc in both cases
    is due to the fact that $m_2 >> m_1$.}
  \label{fig.hatp13ph}
\end{figure}

We now utilize the linear secular theory {\it without} tides developed
in Section~\ref{sec.stable}, and show that the model
predicts the current apsidal state of HAT-P-13 reasonably well.  In
Figure~\ref{fig.hatp13ph}, we compare the apsidal state of HAT-P-13
estimated by secular theory with that obtained from a direct N-body
simulation done by the HNBody code \citep{rauch2002dda}.  Since
HAT-P-13c's large eccentricity of $\sim0.66$ violates the assumption
of the linear secular theory, it is understandable that the
discrepancy between the two integrations is $\sim 33\%$
(Figure~\ref{fig.hatp13ph}).  It appears that the system librates with a
large amplitude about $\Delta \varpi=0^{\circ}$ so that the system is
not far from the aligned separatrix $s^+$ (Figure~\ref{fig.ssphasep}).
Here, we do not take account of any additional apsidal precessions.
By taking account of the GR effect, we find that the results stay
similar --- the system librates with a large amplitude.  However, by
considering all of the additional apsidal precessions, the system appears
to circulate.  Thus, HAT-P-13 is likely to be yet another example of a
multi-planet system being near a secular separatrix
\citep{barnes2006478}.  The near-separatrix state of this system is
somewhat surprising given that the tidal decay time is a tiny fraction
of the stellar age (Table~\ref{tab_orb_el}) and so the system should
have long ago damped to the apsidally-locked state. Perhaps, given the
uncertainty in $Q^{\prime}$, our estimation of $\tau_e$ is off by a
large factor. Alternatively, there might be a third as-yet-undiscovered
planet affecting this secular system.
% Soko - I don't understand this - OK to omit?
%Alternatively, HAT-P-13 is an example where the secular interactions
%between two observed planets are slowing down the circularization
%significantly.

\begin{figure*}%[p]
  \begin{tabular}{cc}
    \includegraphics[width=0.45\textwidth]{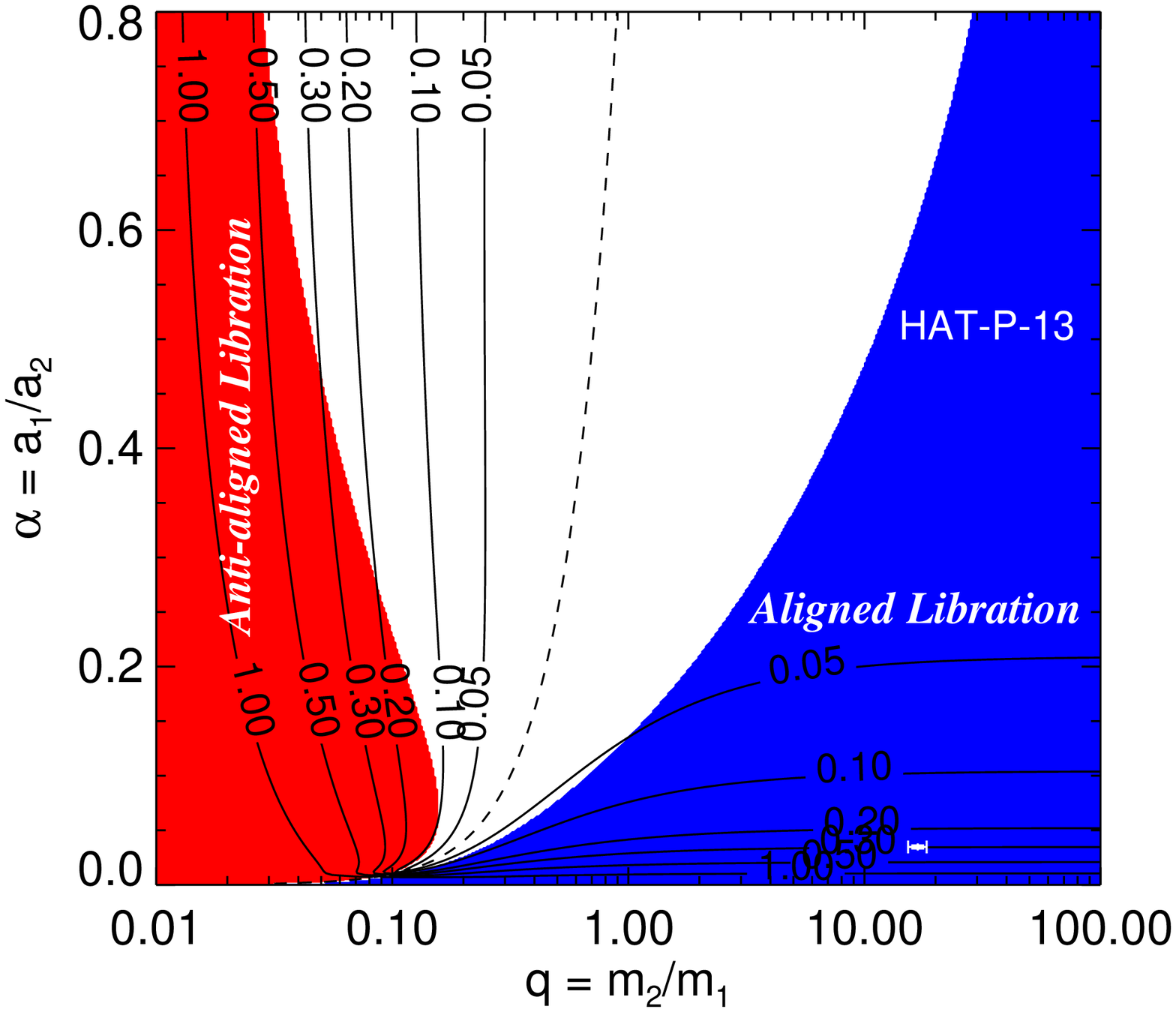} &
    \includegraphics[width=0.45\textwidth]{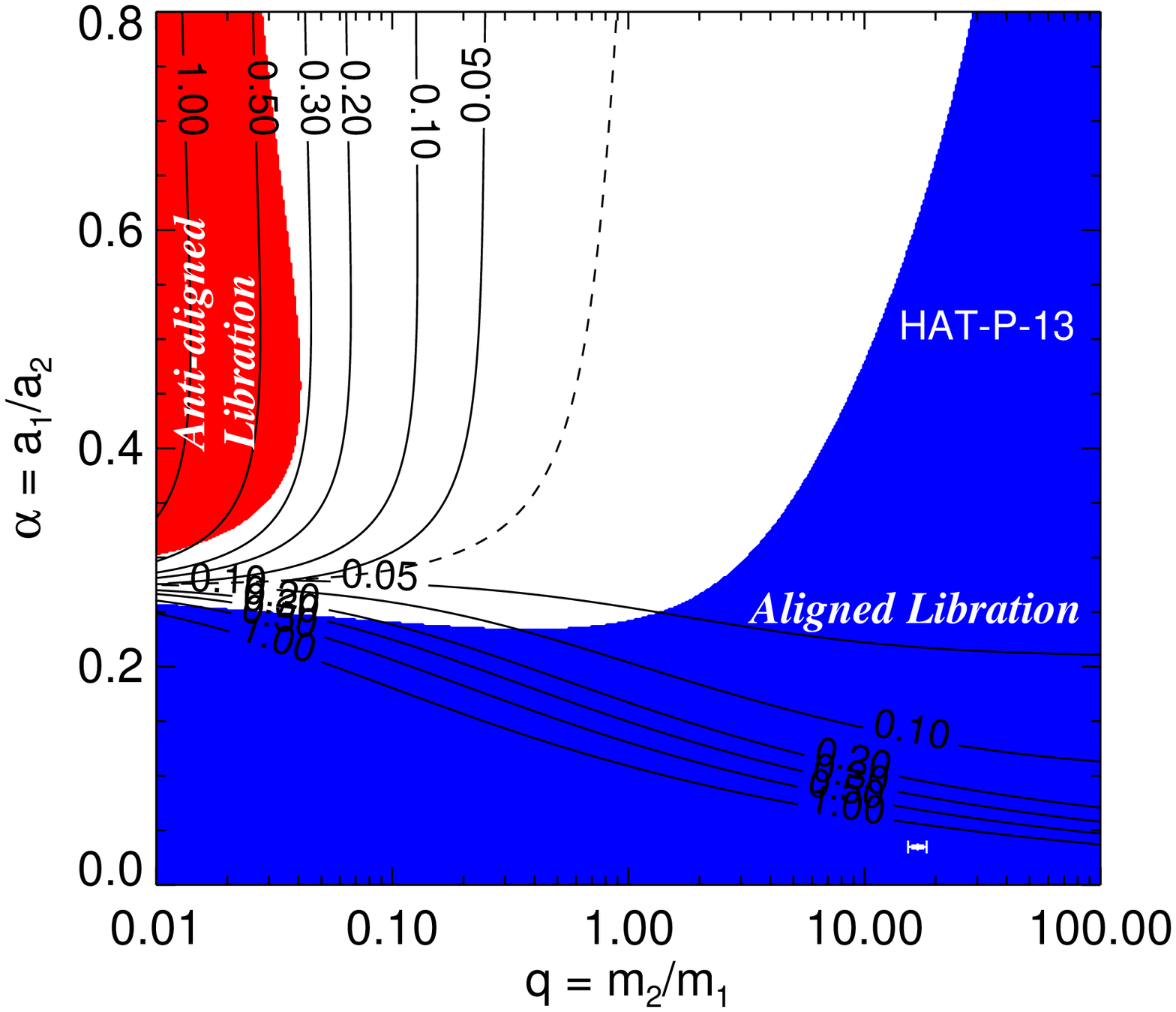} \\    
    (a) without additional precessions & (b) with General Relativity\\
    \includegraphics[width=0.45\textwidth]{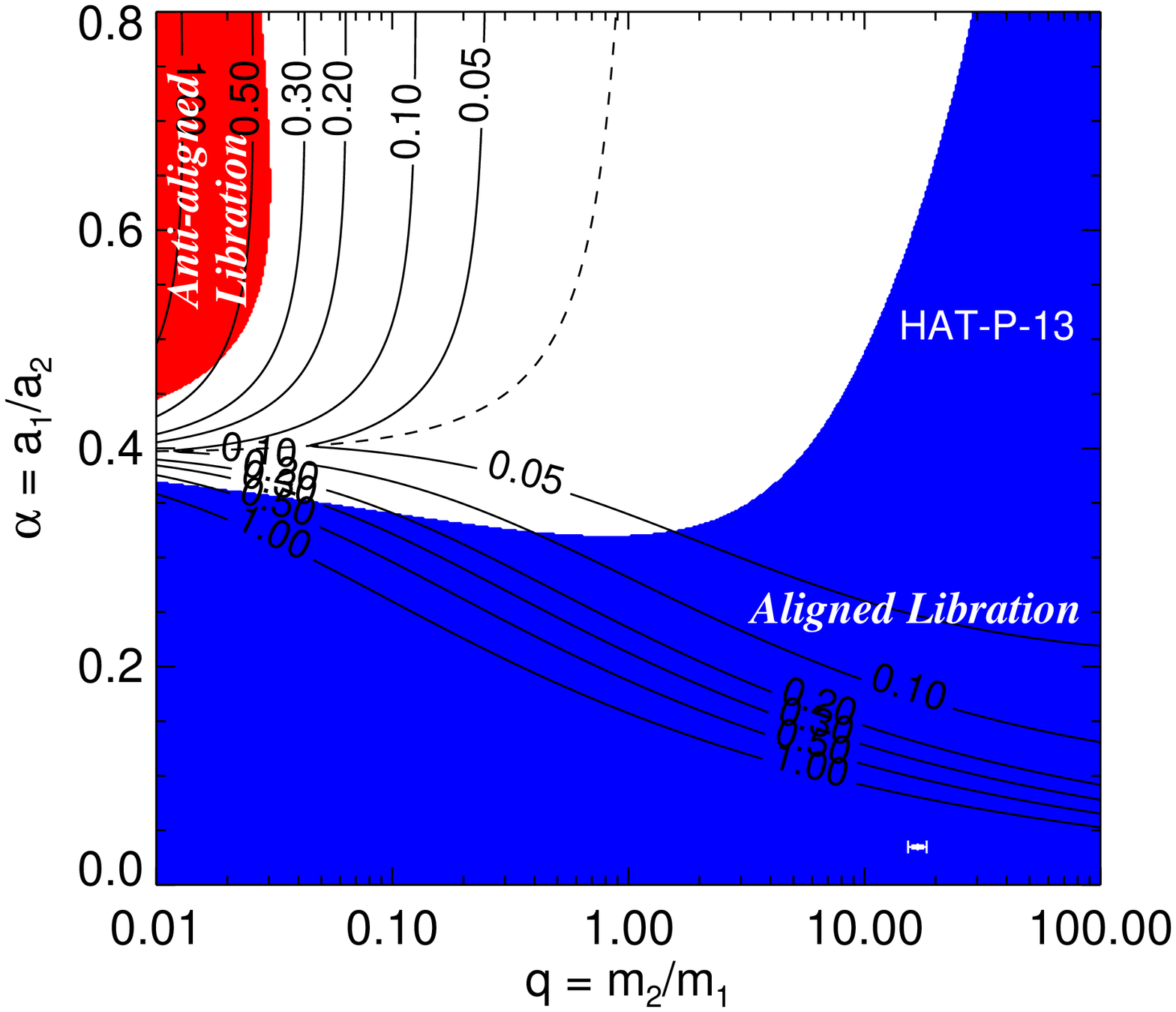} &
    \includegraphics[width=0.45\textwidth]{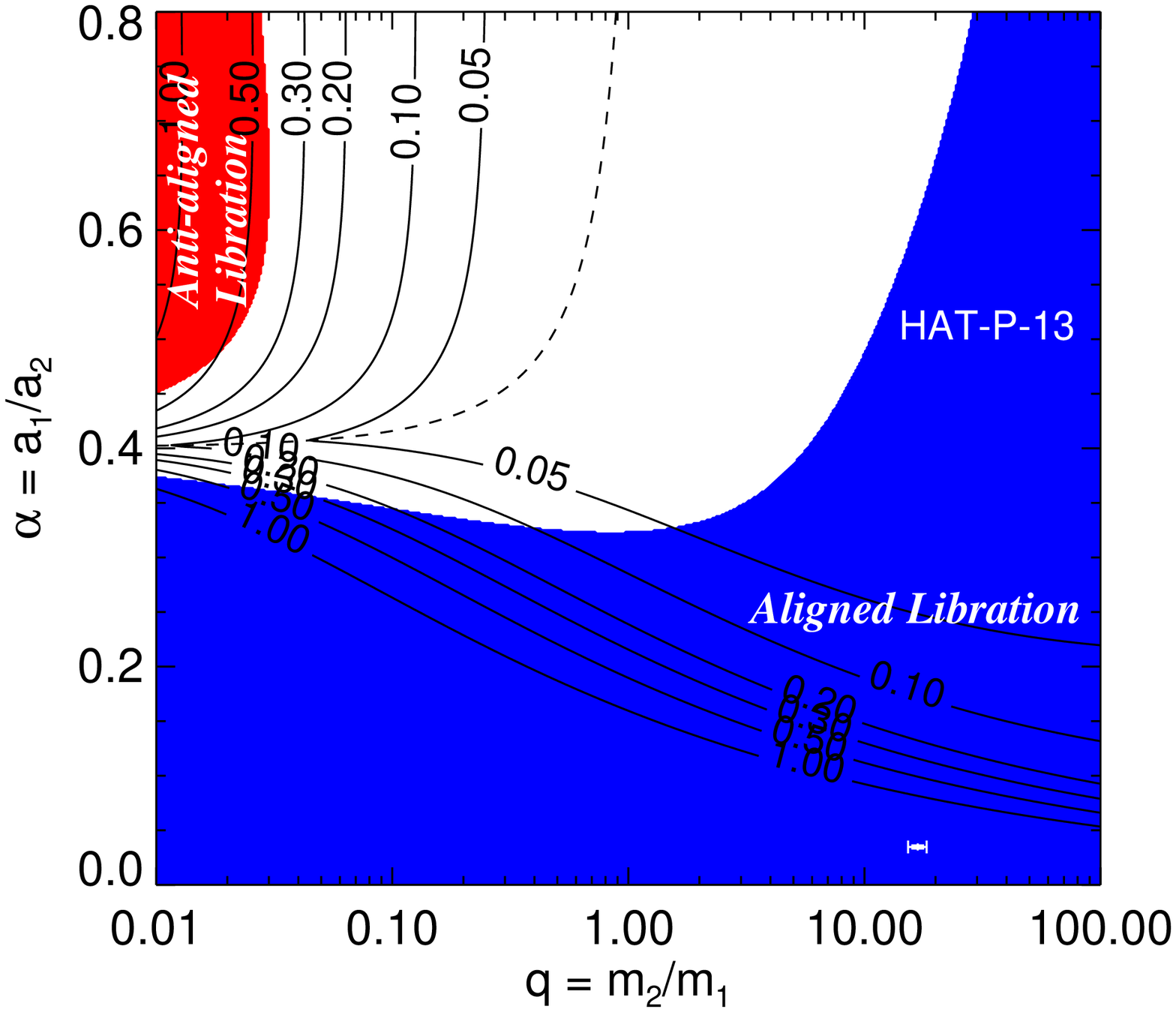} \\
    (c) with GR and planetary tides & (d) with all additional precessions\\
  \end{tabular}
  \caption[Orbital states of possible companions for HIP 14810b]{
    Orbital states of possible companions for HAT-P-13b (a) with no
    additional apsidal precessions, (b) with GR precession, (c) with
    GR and planetary tidal precessions, and (d) with different assumptions 
    about the orbital precession terms from Section~\ref{sec.gr}.  Each point in
    these $q$ - $\alpha$ plots represents an outer companion with
    corresponding mass and semi-major axis.  The dashed curve divides
    the plane into regions in which the aligned mode damps faster
    (top left) and the anti-aligned mode damps faster (bottom right).
    Furthermore, in the shaded regions, either the aligned mode (blue
    shading) or the anti-aligned mode (red shading) can survive tidal
    dissipation and last longer than the age of the system
    ($5\,$Gyr).  Last, the solid contour lines represent the
    eccentricity of the outer planet assuming that it is in the
    long-lived mode.  The location of HAT-P-13c is marked by a white
    symbol with error bars. }
  \label{fig.hatp13}
\end{figure*}

We next compare the current state of the system with the expectation
from tidal-damping theory as developed in Section~\ref{sec.damping}.
We assume that the parameters of the inner planet's orbit are known,
assume further that one mode has fully damped away (despite the
contrary evidence of Figure~\ref{fig.hatp13ph}), and proceed to predict
parameters of the outer planet. We begin by asking which mode is
favored, which depends on the damping rates given by
Equation~(\ref{eq.gamma}), or Equation~(\ref{eq.gammagr}) when other apsidal precession effects are
important.

When a system is locked into one of the secular eigen-modes, the outer
and inner orbits have a predictable eccentricity ratio
$|\eta_{\pm}|=(e_2/e_1)_\pm$ (see Equations~(\ref{eq.eta}) and
(\ref{eq.etagr})).  In Figure~\ref{fig.hatp13}, we show contour plots of
the predicted eccentricity of the outer planet $e_2$ for the more
slowly damped eigen-mode in the parameter space ($q$, $\alpha$).  The
dashed curve divides the space into a region in which the slow
anti-aligned mode persists (top left, red area) and a region in which the
slow aligned mode survives (bottom right, blue area).  In these shaded
areas, the lifetime of the slower mode is longer than the age of the
system. Conversely, in the white central area, both modes should have
already damped away and both orbits would be circular by now. Given
the inner planet's non-zero $e_1=0.013$ (Table~\ref{tab_orb_el}) and 
that $\tau_e <<\tau_{\rm Age}$, we expect the outer planet to
be in one of the shaded regions. Furthermore, since $e_2 > 1$ correspond to
unbound orbits, these parts of the shaded areas in
Figure~\ref{fig.hatp13} are also off limits.

The boundaries of these colored areas are determined by equating the
age of the system (5 Gyr for HAT-P-13) to the circularization time $\tau_e$ 
in Table~\ref{tab_orb_el}.
%the longer of the two damping timescales $1/\gamma_\pm$ (see Fig.~\ref{fig.modedamp}). 
An older system age $\tau_{\rm Age}$ and/or faster damping timescale
$\tau_e$ would expand the white area outward away from the dashed line. 
%Furthermore, depending on initial conditions, it may take
%several damping timescales to reach the apsially-locked state
%\citep[e.g.,][]{mardling20071768,batygin200923}. 
If tides have not been active over the full stellar age,
as is possible for a recent resonance crossing or a planet-planet
scattering event, the white area would shrink inward toward the dashed
line.

The effect of the additional apsidal precessions is substantial
and can be quantified by comparing panels (a)-(d) in Figure~\ref{fig.hatp13}.
The panels (a)-(d) show the estimates from the linear secular theory (a) without additional 
precessions, (b) with GR precession, (c) with GR and planetary tidal precessions, and (d) 
with all apsidal precessions, respectively.
By comparing panel (a) with panels (b) and (c), we find that GR and planetary tidal 
precessions significantly change the $q-\alpha$ plot.
Conversely, from the comparison of panels (c) and (d), we can tell that the other 
precessions have negligible effects on HAT-P-13.
%These perturbations are large for close-in planets and can
%dominate planetary perturbations when the outer companion is less
%massive (small $q$) and/or distant (small $\alpha$). 
%However, the relative strengths of these effects differ from one another.

\begin{figure}%[p]
  \plotone{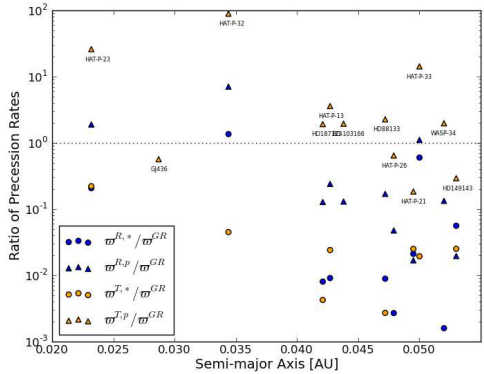}
  \caption{Comparisons of apsidal precession rates for planetary systems listed in 
  Table~\ref{tab_orb_el}. Blue and orange symbols represent the ratios of rotational and GR 
  precessions ($\varpi^{R}/\varpi^{GR}$) and tidal and GR precessions 
  ($\varpi^{T}/\varpi^{GR}$), respectively. 
  Circles and triangles correspond to precessions due to stellar and planetary deformations, respectively. 
  For all of the close-in systems listed here, either GR or planetary tidal precession dominates 
  the additional apsidal precession.}
  \label{fig.precessions}
\end{figure}

In Figure~\ref{fig.precessions}, we compare the GR precession rate with each of other 
precession rates for all of the systems listed in Table~\ref{tab_orb_el}. 
For these close-in systems, we find that either GR or 
planetary tidal precession dominates the additional apsidal precession, while 
the effects of stellar tidal and rotational precessions tend to be much smaller.
For HAT-P-13, the precession rate due to tidal deformation of the inner 
planet is a factor of a few larger than the GR precession rate, while the other 
precessions are much smaller than GR.
 
As discussed in Section~\ref{sec.gr}, adding extra precessions diminishes 
the anti-aligned area (red shading) due to faster damping rates but significantly 
expands the aligned area (blue shading) in accordance with Equation~(\ref{eq.gammagr}).
Notice that, as expected, the lower left quadrant of the plot (i.e., small
$q$ and small $\alpha$) experiences the greatest changes from panel (a) to panels (b)-(d). 
Conversely, the changes to the shading of the other three quadrants are relatively
minor. The dashed lines in Figures~\ref{fig.hatp13}(a) and (b)-(d) are given by
$q = \sqrt\alpha$ and $q + \kappa = \sqrt\alpha(1+\alpha^2\kappa)$,
respectively, where $\kappa$ includes only the corresponding terms in (b)-(d). 
These expressions simply compare the real diagonal elements of the respective $A$ matrices. 
The anti-aligned mode damps most quickly if the inner planet precesses faster while the aligned
mode damps first if the outer planet precesses faster.  Along this
dashed line, the difference between the mode precession rates
$(g_+^s - g^s_-)$ is minimized and the mode damping rates $\gamma_\pm
=(\lambda_1+\lambda_2)/2$ are identical.

Last, note that including additional apsidal precessions significantly changes the contours for
the outer planet's eccentricity. In the region of aligned libration,
the eccentricity contours are moved upward by the inclusion of these precessions,
indicating that a more eccentric outer planet is required to maintain
the mode, for a given $\alpha$ and $q$.  
In the anti-aligned region, the contours move to the left, indicating that a
lower eccentricity on the outer planet is needed to preserve the mode.
The reasons for these changes were discussed in Section~\ref{sec.gr}. 

The actual location of the outer planet HAT-P-13c is marked in all of the 
panels with white error bars. 
%notice first that the different predictions for $e_2$ indicate that these additional 
%precessions cannot be neglected for this system. 
The planet resides well within the more slowly damped aligned
mode region, as expected from linear secular theory (see also
Figure~\ref{fig.hatp13ph}).
%There
%is a discrepancy between the observed outer planet HAT-P-13\,c has an
%eccentricity of $0.662\pm0.054$ while Figure~\ref{fig.hatp13}b suggest
%that $e_2$ should be about 1.0. 
Figures~\ref{fig.hatp13}(b)-(d) suggest that the eccentricity of the
outer planet exceeds one, while the observed value is actually
$0.662\pm0.054$ (Table~\ref{tab_orb_el}).  It is clear that $e_2 \ge 1$ is an
unphysical result, which might be attributable to: 
(1) using linear secular theory despite large eccentricities, and 
(2) assuming that one mode dominates despite the evidence from
Figure~\ref{fig.hatp13ph}. 
%iii) assuming the minimum masses for
%the planets from Table~\ref{tab_orb_el}. 
Although the quantitative
agreement is not very good, the figure does predict apsidal alignment and
a large eccentricity for the outer planet.
%In fact, our theory, derived for small eccentricities only, does surprisingly well outside
%of its realm of strict validity. 
We accordingly conclude that the secular perturbations between the two
planets in the HAT-P-13 system might be responsible for the
non-zero eccentricity of the inner planet.

This section shows the power and pitfalls of our method. If                   
estimates for the stellar ages and tidal damping timescales are                        
accurate, we can determine whether a given system should                        
currently be near a single eigenmode. The HAT-P-13 system with a                       
stellar age 40 times longer than the estimated damping time should                     
have had ample time to reach such a state, and yet Figure~10 shows that                  
it has not. Perhaps the time estimates are inaccurate or perhaps there                 
was a recent disruptive event in the system. In either case, this                      
suggests that a certain amount of caution is warranted when proceeding                 
to investigate single-planet systems. 
With this in mind, in the following two sections, we study single-planet systems with and
without observed linear trends in the stellar radical velocity that
might be indicative of a companion. We investigate whether the observed
non-zero eccentricities could be explained by unseen potential
companions.

\subsection{Single-planet Systems with a Hint of a Companion} \label{sec.1ptrend}

%\subsection{Constraints on Possible Companions of Close-in Planets} \label{sec.1plsystems}

As shown in Sections~\ref{sec.damping} and \ref{sec.apsidal}, a
two-planet system should have evolved into either an aligned or an
anti-aligned apsidally locked state when the tidal dissipation is
strong enough.  Equations~(\ref{eq.etasgr}) and (\ref{eq.gammagr})
thus provide a single constraint on the three parameters of the
unknown outer companion: the mass ratio $q$, the semi-major axis ratio
$\alpha$, and the eccentricity ratio of the two planets.  Therefore,
we can predict a range of possible companions that might force a
non-zero eccentricity on an observed close-in planet.  We illustrate
our method with several examples here and in the next section.

There are 16 single, close-in planet systems with $P_{\rm
  orb}\leq20\,$days and non-zero eccentricities that have an observed
linear trend in the stellar RV, which indicates the
possible existence of a companion on a more distant orbit.  For these
systems, we can place a unique constraint on the potential companion.
For simplicity, we exclude the two systems that have a large projected
stellar obliquity (HAT-P-11 and WASP-8), and compare the estimated
$\tau_e$ with the stellar age $\tau_{\rm Age}$ for each of the
remaining systems, as described earlier.  We find that 5 of 14 
systems (GJ\,436, BD\,-10\,3166, HAT-P-26, WASP-34, and HD\,149143)
have $\tau_e < \tau_{\rm Age}$. We list parameters for these systems
in the middle section of Table~\ref{tab_orb_el}. Because the properties of
the putative companions are unknown, we cannot test for a strong
secular interaction as in the previous section and so we investigate
all five systems.
 
\begin{figure*}%[p]
  \begin{tabular}{cc}
    \includegraphics[width=0.48\textwidth]{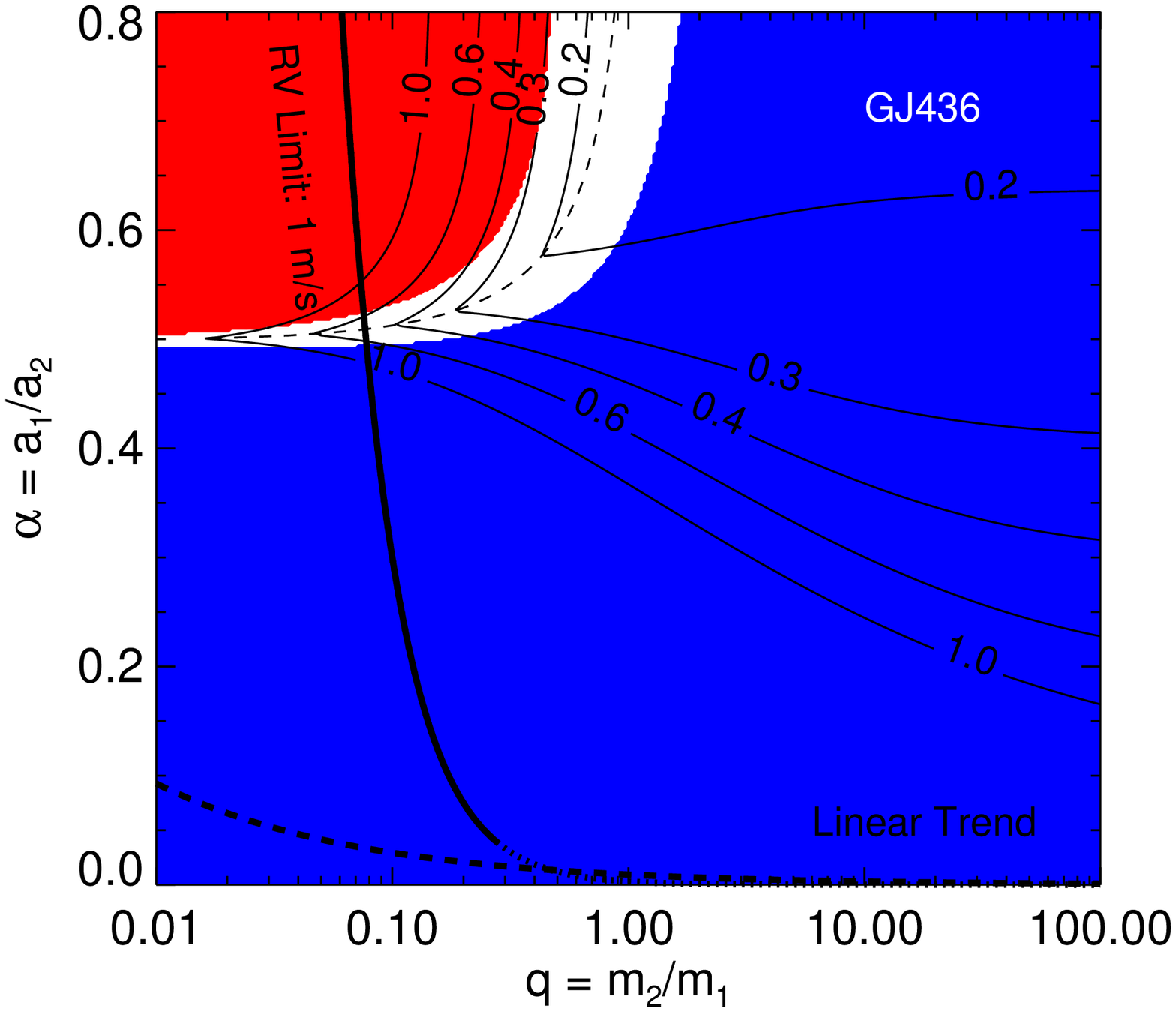} &
    \includegraphics[width=0.48\textwidth]{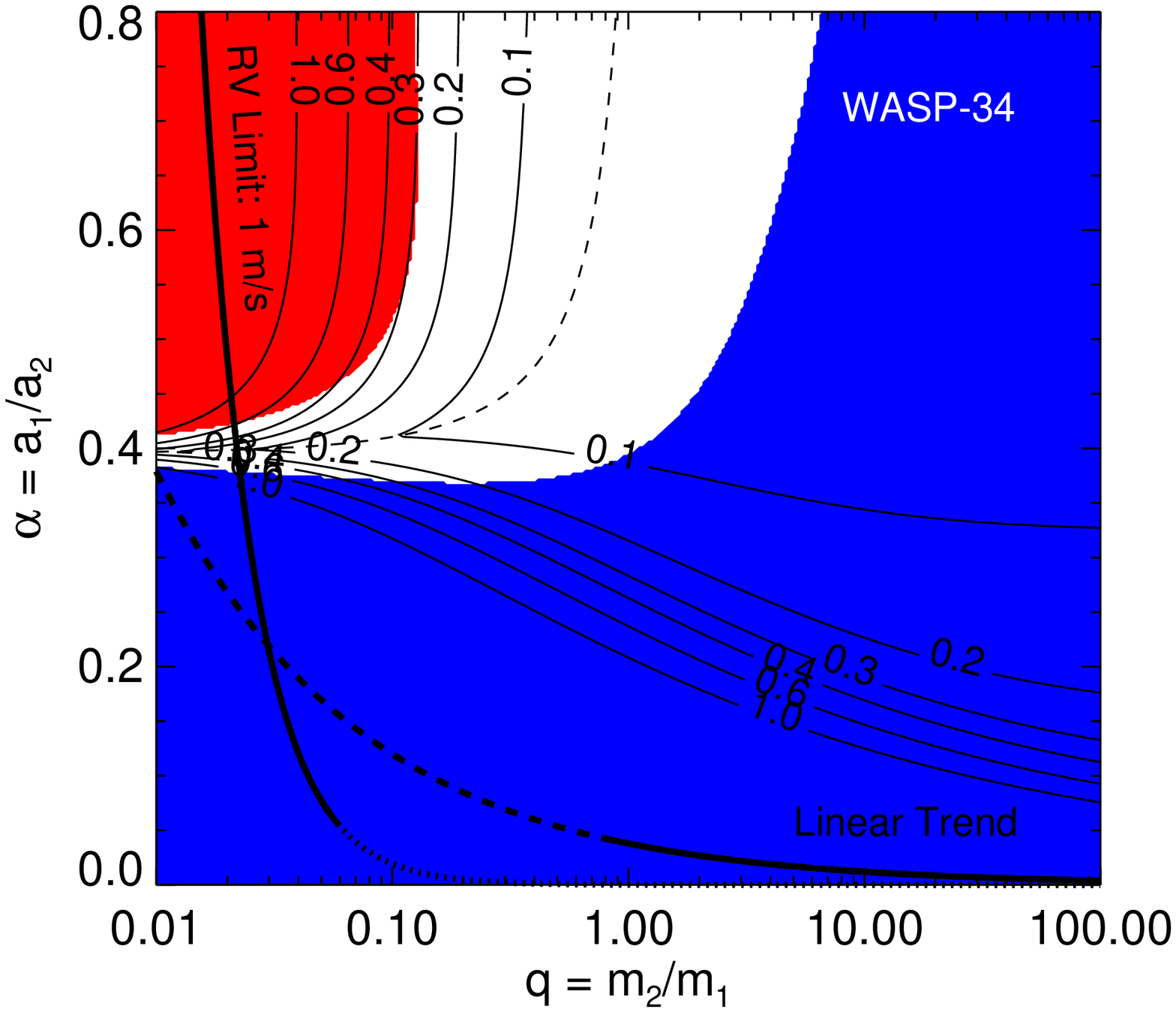} \\
    (a) & (b)\\
  \end{tabular}
%  \plottwo{plots/GJ436_e2gr_tiderot.eps}{plots/WASP-34_e2gr_tiderot.eps}
  \caption[Eccentricities of possible companions for ``hot-Neptune''
  GJ 436b and GJ 674b ]{Eccentricities of possible companions for hot
    Jupiters (a) GJ\,436, and (b) WASP-34 including all apsidal
    precession effects as in Figure~\ref{fig.hatp13}(d).  In the white
    area, both modes should damp away within the age of the system of
    6 and 6.7 Gyr, respectively. The central stars of these systems
    each have an observed linear trend in their radial velocities,
    1.36 and 55 ${\rm m \, s^{-1} \, yr^{-1}}$, respectively, that might 
    be indicative of second planets. 
    The thick, dashed curves represent these constraints. For WASP-34, the region
    to the right of the solid portion of the curve represents where a potential
    candidate is expected to exist \citep{smalley2011130}.  
    The dotted curves represent the RV observation limit of $1\,{\rm
      m\,s^{-1}}$; only planets to the right of the nearly vertical solid part of this
    curve are detectable with current technology in a 1 yr observational period.  }
  \label{fig.e2lt}
\end{figure*}

Figure~\ref{fig.e2lt}(a) is similar to Figure~\ref{fig.hatp13}(d), but for
the planetary system around GJ\,436. The thick, dashed curve is an
upper limit to the outer planet's mass estimated from the observed
linear trend. Here, we simply assume that the minimum mass of a
potential outer planet is expressed as $m_{2}=a_2^2 \, a_{lt}/G$,
where $a_{lt}=1.36\pm0.4\,{\rm m\,s^{-1}yr^{-1}}$ is an observed
linear trend \citep{maness200790}.\footnote{This long-term trend has
  not been confirmed by HARPS \citep{bonfils201111115019}.}  The
dotted curve indicates an observation limit for the RV method.  
The solid portion of this curve is plotted as a
reference, and it shows the limit estimated for a 1 yr observation period.  
To the right of this solid curve, a full orbit of 
a hypothetical outer planet is observable within a year.  To plot
this, we express the mass of a potential outer planet as $m_2\sin i =
m_* \, v_*\sin i /\sqrt{Gm_*/a_2}$, where $i$ is the viewing angle,
and assume the RV limit of $v_*\sin i=1\,{\rm m\,s^{-1}}$.
%the M dwarf star GJ 436; 
%here we plot contours of the companion's eccentricity rather than the eccentricity ratios.  
%With only 1.3 Neptune masses, GJ 436b is the first
%Neptune-sized extrasolar planet detected through a radial velocity
%survey \citep{butler2004580}.  It has recently been found to transit
%its host star \citep{gillon200713}, which removes the uncertainty
%in the determination of its mass.  The planet is $0.023\,AU$ away from
%the star (orbital period $\sim2.6$ days), and has an orbital
%eccentricity of about 0.16.  

The observed planet GJ\,436\,b is about a Neptune-mass object ($m_1\sin
i=0.073\,m_J \sim 1.35\,m_N$) which is $0.0287\,$AU away from the
central star (orbital period $\sim2.6\,$days), and has an orbital
eccentricity of $0.16\pm0.019$ (Table~\ref{tab_orb_el}).  If this
eccentricity is due to another planet and the system has damped to an
eigen-mode, then eccentricity contours in Figure~\ref{fig.e2lt}(a) show
that broad aligned and anti-aligned regions are allowed for a
potential companion, except for small ($m_2\lesssim 0.15\,m_1 \sim 0.2\,m_N$) and/or
distant ($a_2\gtrsim 6.3\,a_1 \sim 0.18\,$AU) planets.
%ruled out by the observed linear trend.  
Furthermore, the plotted RV limit indicates that nearly all hypothetical outer planets 
which could be responsible for the high eccentricity of GJ\,436\,b should be observable 
within a year.
%This is because the circularization time ($\tau_e\sim1\,{\rm Gyr}$) is shorter than, 
%but comparable to the stellar age ($\tau_{\rm Age}\sim4.18\,{\rm Gyr}$).
%excludes almost the entire anti-aligned region and
%all companions that are less that $1/10$ as massive as GJ 436b, or
%more than 5 times further away -- these planets are simply too weak to
%force the inner planet's eccentricity.  In fact, if we consider the
%age of the system, which is estimated to be older than 3 billion years
%\citep{gillon200713} and probably much older, most possible outer
%companions are much more massive than GJ 436b.  Larger planets in this
%region are very likely within the radial-velocity detection window,
%but none have been found.  Most planets outside this region, however,
%cannot be observationally excluded.  
The curve representing the maximum linear trend, however, is far below
the $e_2=1$ contour, implying that this potential planet cannot be
responsible for the current eccentricity of the observed planet; given
its great distance, the secular interactions are simply too weak.
This result is consistent with the comparison of the secular and tidal
circularization timescales by \cite{matsumura200829}.  Since the
system has a tidal dissipation timescale ($\tau_e\sim2\,{\rm Gyr}$)
comparable to the stellar age ($\tau_{\rm Age}\sim6^{+4}_{-5}\,{\rm
  Gyr}$), a non-zero eccentricity of this planet might also be
explained within the uncertainties of the stellar age.

%Another recently discovered ``hot-Neptune'', GJ 674b in shown in
Another example is shown in Figure~\ref{fig.e2lt}(b) for WASP-34, which
has an observed planet of $m_1\sin i=0.58\,m_J$, $a_1=0.052\,$AU, and
$e_1=0.038\pm0.012$ (see Table~\ref{tab_orb_el}).  Again, most of the
parameter space of the $q-\alpha$ plane is available for a possible
secular companion, except for small ($m_2\lesssim 7.4\,m_E$) and/or
distant ($a_2\gtrsim 0.74\,$AU) planets.
The solid portion of the RV limit again indicates that such a companion should be 
observable within a year.
%due to a relatively long eccentricity damping timescale ($\tau_e\sim1\,{\rm Gyr} < 
%\tau_{\rm Age}\sim6.7^{+6.9}_{-4.5}\,{\rm Gyr}$).
The system has an observed linear trend of $55\pm4\,{\rm
  m\,s^{-1}yr^{-1}}$ \citep{smalley2011130}.  Since the long-term
trend has not reached its maxima or minima, the orbital period of the
outer body has to be greater than twice the RV data baseline.  
%This corresponds to $a_2\gtrsim1.2\,$AU and $m_2\gtrsim0.45\,m_J$
%\citep{smalley2011130}, and is indicated as a thick solid curve in the
%figure.  
The solid portion of the linear trend corresponds to this limit of 
$a_2\gtrsim1.2\,$AU and $m_2\gtrsim0.45\,m_J$ \citep{smalley2011130}.
A yet-to-be-observed companion should lie in the triangular area right of the solid part of 
the linear trend and below the observable region of the RV Limit. 
%Similar to GJ\,436, the linear trend lies in the region where the expected eccentricity of 
%the outer planet exceeds one, but much closer to the critical $e_2=1$
%contour than for GJ\,436.  
It is clear that such a region does not have any overlap with the critical $e_2=1$ contour.
However, they lie relatively close to each other so that 
the uncertainties in both observations and the secular model could bring them closer to have an overlap.  
Our model predicts that the companion planet
responsible for both this trend and the eccentricity of the WASP-34\,b
would have a very high eccentricity.
%This system is very similar to the GJ 436 system,
%except that the star is much younger.  With an age of only about half
%a billion years, most of the parameter space in the $q-\alpha$ plane is available for a possible secular companion.    
%The slightly larger eccentricity of GJ 674b, however, pushes the eccentricity contours
%towards the upper-right in Fig.~\ref{fig.e2}b, which means that all
%possible companions with small masses or large semi-major axes are
%excluded.
%Thus, although there is an overlap between eccentricity contours and
%the thick dashed curve, a potential outer planet cannot be responsible
%for the current eccentricity of WASP-34\,b.  

If no companion is present, how do we explain WASP-34? For this
system, although the estimated eccentricity damping time 
($\tau_e\sim700\,{\rm Myr}$) is short compared with the stellar age 
($\tau_{\rm  Age}\sim6.7^{+6.9}_{-4.5}\,{\rm Gyr}$), the eccentric orbit model 
gives only a slightly better fit than the circular one \citep{smalley2011130}. 
Thus, the inner orbit might well be circular. 
Alternatively, if the orbit is truly eccentric, we would
need to assume about an order of magnitude less efficient tidal
dissipation to explain this system.  Last, the system could also 
have undergone some dynamical event lately which changed the original
eccentricities.

The other systems with a linear trend (BD\,-10\,3166, HD\,149143, and
HAT-P-26) show a similar result to GJ\,436 (Figure~\ref{fig.e2lt}(a)), and
thus a potential planet is too far to force the eccentricity of the
inner planet to its current value.  It is interesting that the observed 
eccentricities are low and consistent with zero for BD\,-10\,3166 and
HD\,149143 \citep{butler2000504,fischer20061094}, and poorly
constrained for HAT-P-26 \citep{hartman2011138}.  BD\,-10\,3166 has a
short tidal dissipation timescale ($\tau_e\sim420\,{\rm Myr}$)
compared with the stellar age ($\tau_{\rm Age}\sim4.18\,{\rm Gyr}$), so
the circular orbit assumption makes sense. 
HD\,149143 and HAT-P-26 have relatively long dissipation timescales
($\tau_e\sim2.75\,{\rm Gyr}$ and $\sim2.6\,{\rm Gyr}$, respectively)
compared with the stellar ages ($\tau_{\rm Age}\sim7.6\pm1.2\,{\rm Gyr}$
and $\sim9^{+3}_{-4.9}\,{\rm Gyr}$, respectively).  Thus, uncertainties
in the age estimates and/or in the tidal dissipation rates could allow
close-in planets to maintain their eccentricities without assistance.

In summary, we have found no single-planet systems, where the none-zero eccentricities 
could be explained by perturbations from hypothetical planets corresponding to 
the observed linear trends. 
However, their orbits could be circular (WASP-34, BD\,-10\,3166, HD\,149143, and
HAT-P-26), or the eccentric orbit could be explained within the uncertainties in the estimated stellar age (GJ\,436).
%
%($\tau_e\sim4\,{\rm Gyr} < \tau_{\rm Age}\sim6^{+4}_{-5}\,{\rm Gyr}$ for GJ\,436, and 
%$\tau_e\sim6\,{\rm Gyr} < \tau_{\rm Age}\sim7.6\pm1.2\,{\rm Gyr}$ for HD\,149143).  
%Thus, their non-zero eccentricities could be explained simply within the stellar age's uncertainties.
%%% The following paragraph is WRONG.
%Finally, HAT-P-26 shows a similar trend to WASP-34 (Fig.~\ref{fig.e2lt}b), with an overlap between the
%linear trend curve and the libration regions.  Different from WASP-34,
%the linear trend falls below the detection limit, and thus the current
%observations may not be able to confirm an additional planet for this
%system.
%%% The WRONG paragraph ENDS here.

\subsection{Single-planet Systems with no Hint of a Companion} \label{sec.1pnotrend}

Given that there are many single, close-in planet systems without linear
trends, we focus on planets whose non-zero eccentricities are
hardest to explain --- the closest-in exoplanets.  There are eight single-planet 
systems with an orbital period $P_{\rm orb}\leq 5\,$days,
eccentricity $\geq 0.1$, and a small or unknown stellar obliquity.
Comparing the tidal timescale to the stellar age, we find that five of
eight such systems have $\tau_e < \tau_{\rm Age}$ (HAT-P-21, HAT-P-23,
HAT-P-32, HAT-P-33, and HD\,88133, see Table~\ref{tab_orb_el}).
%Out of these, HAT-P-21 is a hot Neptune system, while the others are hot Jupiter systems.
All of these are hot Jupiter systems.  For the remaining three
systems, the stellar ages of KOI-254 and Kepler-15 are unknown, while
GJ\,674 has a very long $\tau_e>10\,$Gyr compared with the stellar age
($\sim0.55\,$Gyr).

\begin{figure*}%[p]
  \begin{tabular}{cc}
    \includegraphics[width=0.48\textwidth]{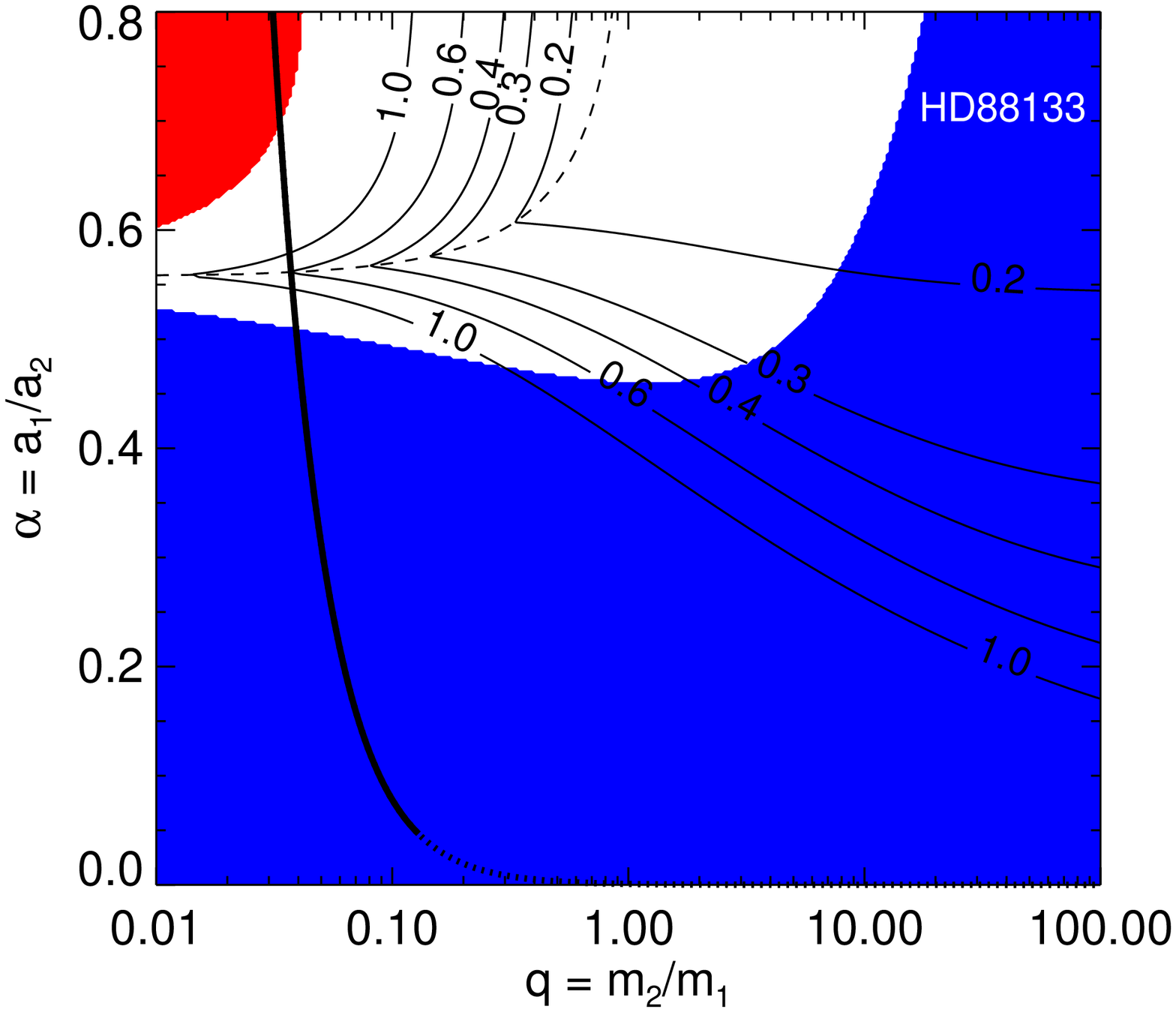} &   
    \includegraphics[width=0.48\textwidth]{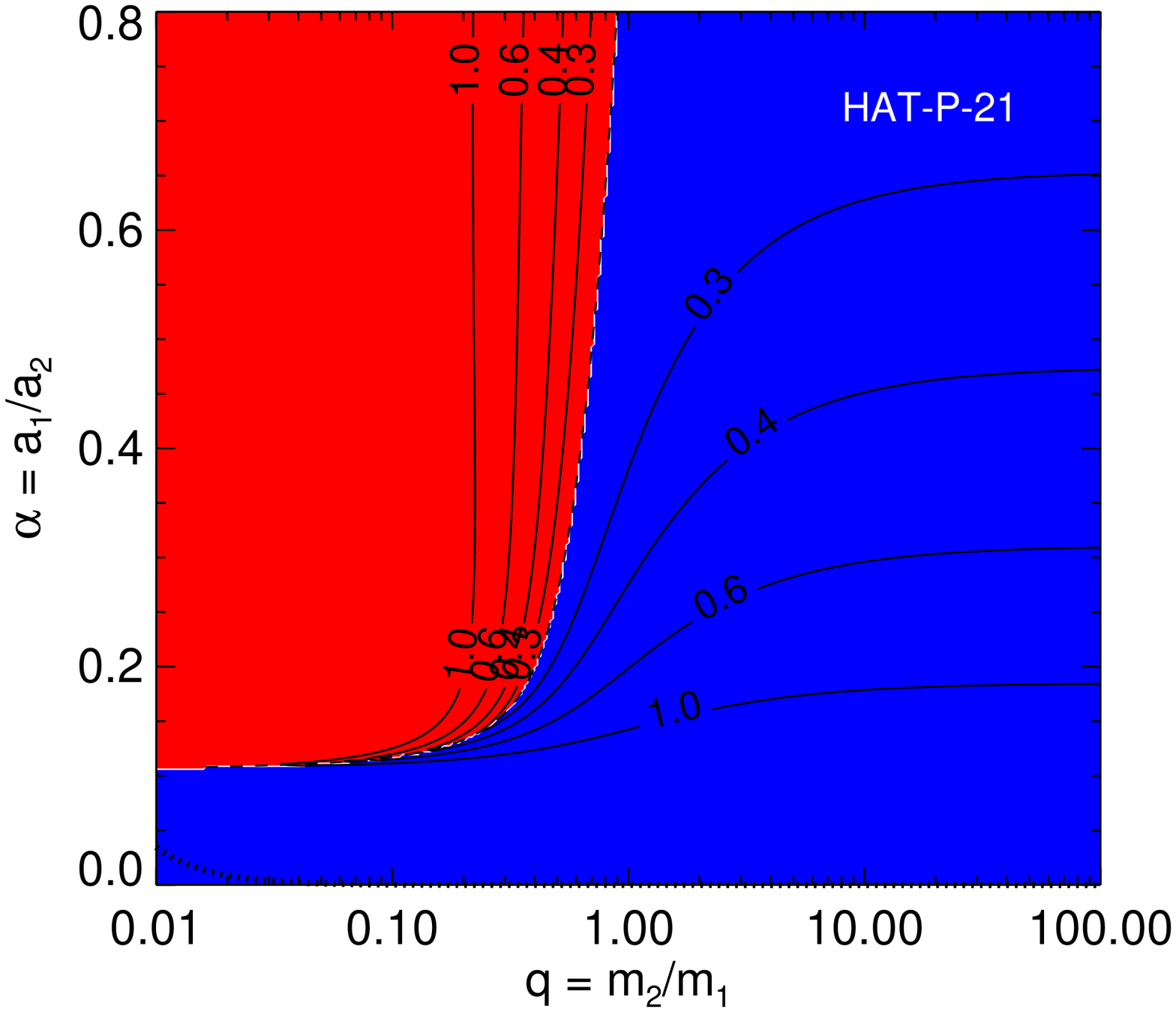} \\  
    (a) & (b) \\
  \end{tabular}
%  \plottwo{plots/HD88133_e2gr_tiderot.eps}{plots/HAT-P-21_e2gr_tiderot.eps}
  \caption[]{Eccentricities of possible companions for hot Jupiters (a)
    HD\,88133, and (b) HAT-P-21 including all apsidal precession
    effects as in Figure~\ref{fig.hatp13}(d).  In the white area, both
    modes should damp away within the age of the system of 9.56 and
    $10.2\pm2.5\,$Gyr, respectively. Because the inner HAT-P-21 planet has
    a long damping time of about 6 Gyr, all eccentricities can be
    sustained and almost no white area is visible. The dotted 
    curves represent the RV observation limit of $1\,{\rm
      m\,s^{-1}}$, and the solid portion indicates a 1 yr observing period. 
    All potential HD\,88133 outer planets reside in
    the aligned zone are detectable. Aligned and anti-aligned
    solutions exist for HAT-P-21, and nearly all outer planets are
    detectable. }
  \label{fig.e2}
\end{figure*}

Figure~\ref{fig.e2}(a) is similar to Figure~\ref{fig.hatp13}(d), but for
the planetary system around HD\,88133.  The planet is $0.0472\,$AU
away from the central star with an orbital period of $3.4\,$days and
an eccentricity of $0.13\pm0.072$ (see Table~\ref{tab_orb_el}).  Since
the circularization time is estimated to be very short
($\tau_e\sim380\,{\rm Myr}$) compared with the stellar age ($\tau_{\rm
  Age}\sim9.56\,{\rm Gyr}$), a moderately high eccentricity of this
planet is surprising.  Figure~\ref{fig.e2}(a) excludes the entire
anti-aligned region for a potential companion.  The figure also
excludes most companions with small mass ($m_2\lesssim 0.03\,m_J$) 
and/or long orbital period ($a_2\gtrsim 0.295\,$AU). This is understandable 
because these planets would have weak secular interactions with the
inner planet.  Thus, a potential companion is expected to be in the
aligned region, massive, and close to the star.  As the RV observation limit shows, 
such massive planets in the aligned
libration region would be observable within a year, although no such
planet has been found.  HD\,88133 has a low stellar jitter
($\sim3.2\,{\rm m\,s^{-1}}$), and the eccentric orbit assumption works
only slightly better than the circular one \citep{fischer2005481}.
Thus, unless the tidal quality factor for this system is very
different from what we have assumed here, we argue that the true
eccentricity of HD\,88133\,b is actually near zero.  Follow-up
observations would better constrain the eccentricity and the
existence or absence of a potential companion for this system.

The case for HAT-P-21 is shown in Figure~\ref{fig.e2}(b).  The planet is
$0.0495\,$AU away from the central star (orbital period
$\sim4.1\,$days), and has an orbital eccentricity of $0.23\pm0.016$
(Table~\ref{tab_orb_el}). Figure~\ref{fig.e2}(b) allows a very broad
parameter space for a possible secular companion; the white 
zone indicating efficient eccentricity damping is nearly absent. 
However, this undercuts our assumption that the system has had time to damp
into a pure eigenmode and, accordingly, the eccentricity contours
are not reliable. 
%If this eccentricity is due to another planet and the system has damped to an eigen mode, Fig.~\ref{fig.e2}a 
%allows a very broad parameter space for a potential companion.
%allows almost the entire aligned and anti-aligned regions for a potential companion.
If we proceed with the dubious assumption of a single mode, the
figure does not allow most companions with $m_2\lesssim0.82\,m_J$
and/or $a_2\gtrsim0.495\,$AU. Furthermore, the $1\,{\rm ms^{-1}}$ RV limit 
observationally precludes almost any planet that can be
significantly coupled to HAT-P-21\,b. Accordingly, we seek another
explanation for the eccentricity of this system; it can be naturally
explained if tidal dissipation were just slightly less efficient than
we have assumed, since the estimated tidal circularization time is
relatively long ($\tau_e\sim2.6\,$Gyr) compared with its stellar age
($\tau_{\rm Age}\sim9^{+3}_{-4.9}\,$Gyr).
%that is less than $\sim1/10$ as massive as HAT-P-33~b, or more than $\sim10$ times further away.
%, since these planets are simply too weak to force the inner planet's eccentricity.  
%In fact, if we consider the
%age of the system, which is estimated to be older than 3 billion years
%\citep{gillon200713} and probably much older, most possible outer
%companions are much more massive than GJ 436b.  
%The observational limit for the radial velocity search of $1\,{\rm m\,s^{-1}}$ indicates that 
%massive planets in the aligned region are very likely within the radial-velocity detection window, although 
%none has been found.
%We should note that HAT-P-33 has high stellar jitters ($\sim55.1\,{\rm m\,s^{-1}}$) \citep{hartman201159}. 
%This means that, at the orbital radius of HAT-P-33\,b, a planet less massive than $\sim0.5M_J$ would not be detected, 
%so there could be a companion which is not discovered.
%However, this high jitter also means a poorly determined orbital eccentricity for HAT-P-33\,b.
%In fact, the circular orbit model is more preferred in terms of the Bayesian Information Criterion \citep{hartman201159}.

The other systems (HAT-P-23, HAT-P-32, and HAT-P-33) have a similar
trend to HD\,88133 (Figure~\ref{fig.e2}(a)), with the entire
anti-aligned region being excluded for a potential companion.  Since
all of their circularization times are more than 1-2 orders of
magnitude shorter than the estimated stellar ages, the moderately-high
eccentricities ($e_1>0.1$) of these planets need to be explained.  From
figures similar to Figure~\ref{fig.e2}, we find that potential
companions for these systems tend to be more massive than the observed
planets and thus are likely to be observable by the RV method.  
However, no companions have been found.  It is interesting 
that all of these systems have high stellar jitters
\citep{bakos2011116,hartman201159}.  Although this could mean that
potential companion planets are difficult to observe, high jitters
also lead to poorly-constrained orbital eccentricities.  
All of these planets can also be fit well with the circular orbit
model. Our model suggests that the circular orbits are probably the most likely solution.
Future observations that yield a more accurate solution for the 
eccentricity are needed. 
% to discuss the efficiency of tidal dissipations and/or
% the architectures of these systems.

For our analysis of exoplanetary systems in this section, we did not
explicitly take account of the effects of uncertainties in orbital or
stellar parameters.  In particular, errors in stellar ages and
eccentricities are often large and might change our results
significantly.  We have tested such effects for all of the systems we
discussed in this section, and found that our conclusions will not
change within the currently estimated uncertainties in parameters.
Also, the assumption of an apsidal lock that we made in
Sections~\ref{sec.1ptrend} and \ref{sec.1pnotrend} might be too strong;
instead, it is possible that not enough time has elapsed for complete damping of
one secular mode.
%It is possible that what we have seen for the HAT-P-13 system
%(i.e., although one mode has been significantly damped, its amplitude
%is not yet near zero) may be true for many systems.  
In this case, for
a close-in planet with known mass, semi-major axis, and eccentricity,
the constraint on $m_2$, $a_2$, and $e_2$ is approximate rather than
exact.
%
%
%
%================================================================================================
% SECTION 4
\section{Discussions and Conclusions} \label{sec.conclusions}

The eccentric orbits of single, close-in planets are generally
circularized on timescales shorter than the stellar ages. 
Close-in planets in multiple-planet systems have much longer
tidal circularization timescales and thus are able to maintain
eccentric orbits for the stellar ages or longer. Given this difference
in tidal circularization times, we might expect a difference in the
eccentricity distributions of close-in planets with and without known
companions. The eccentricities forced by secular perturbations from an
outer planet are typically small, however, and so it is perhaps
not surprising that no such difference has yet been observed.

In this paper, we have explored the possibility that the non-zero
eccentricities of close-in planets are due to observed or hypothetical
planetary companions.  We have provided an intuitive interpretation of
a simple secular evolution model of a coplanar two-planet system that
includes both the effect of the orbital circularization
(Section~\ref{sec.damping}) and apsidal precessions due to GR corrections as well as 
tidal and rotational deformations (Section~\ref{sec.gr}).
%and developed a simple secular theory which allows
%us to estimate the apsidal states and orbital eccentricities of
%coplanar two-planet systems, by including the effects of the orbital
%circularization as well as general relativity corrections.  
We have tested our model by comparing the evolution of apsidal states and orbital 
eccentricities with N-body simulations, and found that the agreement between 
the model and the simulations is very good.
%as seen in Section~\ref{sec.model} and \ref{sec.2plsystems}.  
We have also applied our model to all of the relevant
two-planet systems (Section~\ref{sec.2plsystems}), as well as
single-planet systems with and without a long-term trend in the RV 
to indicate a possible second planet (Sections~\ref{sec.1ptrend} and \ref{sec.1pnotrend}, respectively).
%and studied whether actual/hypothetical
%companions are responsible for the current non-zero eccentricity of
%the inner, observed planet.
The following is a summary of our main results.
\begin{enumerate}

\item In the lowest-order secular theory, the evolution of
  non-dissipative two-planet systems is described by a linear
  combination of two modes characterized by pericenter alignment and
  anti-alignment. Eccentricity damping slightly shifts
  the two normal modes from perfect symmetry, which
  speeds up the precession rate of the aligned mode and slows that of
  the anti-aligned mode (see Section~\ref{sec.damping}).

\item Eccentricity damping affects the two modes at different rates.
  Accordingly, the apsidal state of a two-planet system transitions
  between libration and circulation, and eventually is locked to
  either an aligned or anti-aligned state (see
  Section~\ref{sec.apsidal}). The eccentricity of both planets
  subsequently decays at a very slow rate.

\item GR, tidal, and rotational effects increase the
  precession rates of both aligned and anti-aligned modes. As a
  result, they decrease the aligned-mode damping rate, and increase
  the anti-aligned mode damping rate (see Section~\ref{sec.gr}).
%  The damping rate of the
%  anti-aligned-mode is increased (see Section~\ref{sec.gr}).
 
\end{enumerate}
We confirm results of previous studies
\citep[e.g.,][]{wu20021024,mardling20071768,greenberg20118} and show 
that close-in planets in multiple-planet systems can maintain non-zero
orbital eccentricities substantially longer than can single ones.
%the existence of a companion certainly slows the tidal
%circularization of close-in planets and that 
%Just as we were writing up this manuscript, \cite{laskar2012105} 
%presented a similar secular model for an N-planet system.
%They have taken account of tidal damping, GR corrections, as well as oblateness effects. 
%However, the effects of planetary oblateness become comparable to, or greater than 
%GR effects only for very close-in systems ($\lesssim 0.04\,$AU). 
%
We find, however, that there are currently few two- or one-planet
systems that show signs of secular interactions that are strong
enough to significantly slow tidal circularization.  In
Section~\ref{sec.2plsystems}, we find that only one out of eight
systems (HAT-P-13) shows tentative evidence that secular interactions are
slowing orbital circularization. In Section~\ref{sec.1ptrend}, we
apply our model to 14 single-planet systems with a linear trend and 
find that 5 of 14 have $\tau_e<\tau_{\rm Age}$.  Our secular model
predicts that none of their eccentricities is likely to be affected
by hypothetical planets that could cause the long-term linear trends.
%Out of these five systems, four are unlikely to be affected
%by a hypothetical planet which causes the long-term trend.  
%There is an overlap between the linear trend and apsidal lock regions for WASP-34.  
%Our secular model predicts that a linear-trend planet in WASP-34 would not be responsible 
%for the non-zero eccentricity of the observed planet.  
%
%For HAT-P-26, a linear-trend
%planet could be responsible for the observed moderately high
%eccentricity of the inner planet, but the confirmation of this planet
%may be difficult with the present instruments.  
We have further studied eight very close-in ($P_{\rm
  orb}\leq5\,$days), significantly eccentric ($e\geq0.1$)
single-planet systems in Section~\ref{sec.1pnotrend}.  We find five
of eight systems that cannot be explained by a single-planet
orbital circularization with the conventional tidal quality factors.
Potential companions for all of these systems are massive planets in
the apsidally aligned region and should be observable with current
technology. Since all of the host stars have high stellar jitters, it
is possible that the planetary eccentricities are systematically
overestimated or that outer planets are more difficult to observe than
we have assumed here.

Our model has some limitations and caveats.  We have adopted
the leading-order secular theory for two-planet systems, and in
principle our model cannot be applied to high-eccentricity or high-inclination systems.  
However, the model predicts the general trend of
apsidal states fairly well even for a highly eccentric case (see
Section~\ref{sec.2plsystems}). Moreover, recent observations indicate
that multiple-planet systems tend to be well-aligned
\citep[e.g.,][]{figueira2012139,fabrycky20126328}. 
Nevertheless, it is useful to extend this kind of a study to higher
eccentricities and inclinations and to systems with more than two planets.  
Also, we have ignored the slow decay in semimajor axis due
to tides, which should be weaker than eccentricity damping by a factor
of $e_1^2$; this is consistent with the low $e$ assumption made by
linear secular theory. Nevertheless, previous studies have shown that
the eccentricity of the inner planet does damp faster than that of the
outer planet as a result of inward migration
\citep[e.g.,][]{wu20021024,greenberg20118}.  Another consequence of
different damping rates is that the eccentricity ratio does not remain
constant as the apsidally locked state evolves. Our expression for the
mode-damping rates (Equation~(\ref{eq.gammagr})) is consistent with that of
\cite[][see their Equation~19]{greenberg20118}, in the limit of no
migration.  The tidal and rotational deformations of planets and stars
also change orbital precession rates, and these effects might become
more important than the GR effect for very close-in planets or rapidly
spinning stars.  These effects can be easily added to our model using
the techniques of Section~\ref{sec.gr} \citep[e.g.,][]{laskar2012105}.

Overall, our study indicates that secular interactions slow down the
tidal circularization of the inner planet while speeding up that of
the outer planet. Our survey of likely systems available to us in 2012
indicate that secular interactions might not be a dominant cause for
the currently observed hot, eccentric planets. The lack of close-in
planets with strong secular interactions might be partially explained by
inward orbital decay that is accelerated by non-zero eccentricities
\citep[e.g.,][]{adams20061004}. 
Further research should determine whether the scarcity of compact secular systems will
persist. With improved statistics and precision measurements of
close-in exoplanet eccentricities, it might become possible to find
diagnostic differences in the eccentricity distributions for
single- and multiple-planet systems.
%
%Furthermore, recent observations show that hot Jupiters tend to be singletons, while warm
%Jupiters or hot Neptunes are often found in multiple-planet systems
%\citep[][Jason Steffen, private communication in 2011]{latham201124}.
%These differences may imply a variety of formation mechanisms for
%these planets.  However, independent of their formation paths, if
%secular interactions are playing a significant role in keeping the
%eccentric orbits, and if these planets are not lost, the eccentricity
%distributions may be different for clearly single planet systems and
%for multiple-planet systems.  \cite{moorhead20111} estimated the
%expected eccentricity distribution for Kepler planet candidates, and
%found that there is no significant difference in the eccentricity
%distributions between hot Jupiters and other close-in planets, or
%between single- and multiple-planet systems.  These results are
%inconsistent with those from the radial velocity searches
%\citep[e.g.][]{wright20091084,mayor201111092497}.  The future
%observations would be able to constrain these differences in orbital
%distributions, and tell us a clue about the importance of secular
%interactions for close-in planets.
%%\sedit{Refer to Section 3.2, and also write about HJs vs WJs/HNs.}

%\singlespace

\bibliographystyle{apj}
%\bibliography{extrasolar,obs,planet,books,dyn,temp}
%\bibliography{extrasolar_sedit}
\bibliography{secular}

\begin{thebibliography}{56}
\expandafter\ifx\csname natexlab\endcsname\relax\def\natexlab#1{#1}\fi

\bibitem[{{Adams} \& {Laughlin}(2006{\natexlab{a}})}]{adams2006992}
{Adams}, F.~C., \& {Laughlin}, G. 2006{\natexlab{a}}, \apj, 649, 992, 992

\bibitem[{{Adams} \& {Laughlin}(2006{\natexlab{b}})}]{adams20061004}
---. 2006{\natexlab{b}}, \apj, 649, 1004, 1004

\bibitem[{{Bakos} {et~al.}(2011){Bakos}, {Hartman}, {Torres}, {Latham},
  {Kov{\'a}cs}, {Noyes}, {Fischer}, {Johnson}, {Marcy}, {Howard}, {Kipping},
  {Esquerdo}, {Shporer}, {B{\'e}ky}, {Buchhave}, {Perumpilly}, {Everett},
  {Sasselov}, {Stefanik}, {L{\'a}z{\'a}r}, {Papp}, \&
  {S{\'a}ri}}]{bakos2011116}
{Bakos}, G.~{\'A}., {Hartman}, J., {Torres}, G., {et~al.} 2011, \apj, 742, 116,
  116

\bibitem[{{Barnes} \& {Greenberg}(2006{\natexlab{a}})}]{barnes200653}
{Barnes}, R., \& {Greenberg}, R. 2006{\natexlab{a}}, \apjl, 652, L53, L53

\bibitem[{{Barnes} \& {Greenberg}(2006{\natexlab{b}})}]{barnes2006478}
---. 2006{\natexlab{b}}, \apj, 638, 478, 478

\bibitem[{{Batygin} {et~al.}(2009){Batygin}, {Laughlin}, {Meschiari}, {Rivera},
  {Vogt}, \& {Butler}}]{batygin200923}
{Batygin}, K., {Laughlin}, G., {Meschiari}, S., {et~al.} 2009, \apj, 699, 23,
  23

\bibitem[{{Batygin} \& {Morbidelli}(2013)}]{batygin20131}
{Batygin}, K., \& {Morbidelli}, A. 2013, \aj, 145, 1, 1

\bibitem[{{Beaug{\'e}} {et~al.}(2003){Beaug{\'e}}, {Ferraz-Mello}, \&
  {Michtchenko}}]{beauge20031124}
{Beaug{\'e}}, C., {Ferraz-Mello}, S., \& {Michtchenko}, T.~A. 2003, \apj, 593,
  1124, 1124

\bibitem[{{Bobrov} {et~al.}(1978){Bobrov}, {Vasil'Ev}, {Zharkov}, \&
  {Trubitsyn}}]{bobrov1978489}
{Bobrov}, A.~M., {Vasil'Ev}, P.~P., {Zharkov}, V.~N., \& {Trubitsyn}, V.~P.
  1978, \sovast, 22, 489, 489

\bibitem[{{Bonfils} {et~al.}(2013){Bonfils}, {Delfosse}, {Udry}, {Forveille},
  {Mayor}, {Perrier}, {Bouchy}, {Gillon}, {Lovis}, {Pepe}, {Queloz}, {Santos},
  {S{\'e}gransan}, \& {Bertaux}}]{bonfils201111115019}
{Bonfils}, X., {Delfosse}, X., {Udry}, S., {et~al.} 2013, \aap, 549, A109, A109

\bibitem[{{Brouwer} {et~al.}(1950){Brouwer}, {van Woerkom}, \&
  {Jasper}}]{brouwer1950}
{Brouwer}, D., {van Woerkom}, A., \& {Jasper}, J. 1950

\bibitem[{{Burns}(1977)}]{burns1977113}
{Burns}, J.~A. 1977, in Planetary Satellites, ed. J.~A. {Burns} (Tuscon, AZ,
  USA: Univ. of Arizona Press), 113--156

\bibitem[{{Butler} {et~al.}(1999){Butler}, {Marcy}, {Fischer}, {Brown},
  {Contos}, {Korzennik}, {Nisenson}, \& {Noyes}}]{butler1999916}
{Butler}, R.~P., {Marcy}, G.~W., {Fischer}, D.~A., {et~al.} 1999, \apj, 526,
  916, 916

\bibitem[{{Butler} {et~al.}(2000){Butler}, {Vogt}, {Marcy}, {Fischer}, {Henry},
  \& {Apps}}]{butler2000504}
{Butler}, R.~P., {Vogt}, S.~S., {Marcy}, G.~W., {et~al.} 2000, \apj, 545, 504,
  504

\bibitem[{{Butler} {et~al.}(2006){Butler}, {Wright}, {Marcy}, {Fischer},
  {Vogt}, {Tinney}, {Jones}, {Carter}, {Johnson}, {McCarthy}, \&
  {Penny}}]{butler2006505}
{Butler}, R.~P., {Wright}, J.~T., {Marcy}, G.~W., {et~al.} 2006, \apj, 646,
  505, 505

\bibitem[{{Chiang}(2003)}]{chiang2003465}
{Chiang}, E.~I. 2003, \apj, 584, 465, 465

\bibitem[{{Correia} {et~al.}(2012){Correia}, {Bou{\'e}}, \&
  {Laskar}}]{correia2012L23}
{Correia}, A.~C.~M., {Bou{\'e}}, G., \& {Laskar}, J. 2012, \apjl, 744, L23, L23

\bibitem[{{Danby}(1988)}]{danby1988fcm}
{Danby}, J.~M.~A. 1988

\bibitem[{{Fabrycky} {et~al.}(2012){Fabrycky}, {Lissauer}, {Ragozzine}, {Rowe},
  {Agol}, {Barclay}, {Batalha}, {Borucki}, {Ciardi}, {Ford}, {Geary}, {Holman},
  {Jenkins}, {Li}, {Morehead}, {Shporer}, {Smith}, {Steffen}, \&
  {Still}}]{fabrycky20126328}
{Fabrycky}, D.~C., {Lissauer}, J.~J., {Ragozzine}, D., {et~al.} 2012, ArXiv
  e-prints, arXiv:1202.6328

\bibitem[{{Figueira} {et~al.}(2012){Figueira}, {Marmier}, {Bou{\'e}}, {Lovis},
  {Santos}, {Montalto}, {Udry}, {Pepe}, \& {Mayor}}]{figueira2012139}
{Figueira}, P., {Marmier}, M., {Bou{\'e}}, G., {et~al.} 2012, \aap, 541, A139,
  A139

\bibitem[{{Fischer} {et~al.}(2005){Fischer}, {Laughlin}, {Butler}, {Marcy},
  {Johnson}, {Henry}, {Valenti}, {Vogt}, {Ammons}, {Robinson}, {Spear},
  {Strader}, {Driscoll}, {Fuller}, {Johnson}, {Manrao}, {McCarthy},
  {Mu{\~n}oz}, {Tah}, {Wright}, {Ida}, {Sato}, {Toyota}, \&
  {Minniti}}]{fischer2005481}
{Fischer}, D.~A., {Laughlin}, G., {Butler}, P., {et~al.} 2005, \apj, 620, 481,
  481

\bibitem[{{Fischer} {et~al.}(2006){Fischer}, {Laughlin}, {Marcy}, {Butler},
  {Vogt}, {Johnson}, {Henry}, {McCarthy}, {Ammons}, {Robinson}, {Strader},
  {Valenti}, {McCullough}, {Charbonneau}, {Haislip}, {Knutson}, {Reichart},
  {McGee}, {Monard}, {Wright}, {Ida}, {Sato}, \& {Minniti}}]{fischer20061094}
{Fischer}, D.~A., {Laughlin}, G., {Marcy}, G.~W., {et~al.} 2006, \apj, 637,
  1094, 1094

\bibitem[{{Goldreich}(1963)}]{goldreich1963257}
{Goldreich}, R. 1963, \mnras, 126, 257, 257

\bibitem[{{Greenberg}(1977)}]{greenberg1977157}
{Greenberg}, R. 1977, in Planetary Satellites, ed. J.~A. {Burns} (Tuscon, AZ,
  USA: Univ. of Arizona Press), 157--168

\bibitem[{{Greenberg} \& {Van Laerhoven}(2011)}]{greenberg20118}
{Greenberg}, R., \& {Van Laerhoven}, C. 2011, \apj, 733, 8, 8

\bibitem[{{Hamilton}(1994)}]{hamilton1994221}
{Hamilton}, D.~P. 1994, Icarus, 109, 221, 221

\bibitem[{{Hartman} {et~al.}(2011{\natexlab{a}}){Hartman}, {Bakos}, {Kipping},
  {Torres}, {Kov{\'a}cs}, {Noyes}, {Latham}, {Howard}, {Fischer}, {Johnson},
  {Marcy}, {Isaacson}, {Quinn}, {Buchhave}, {B{\'e}ky}, {Sasselov}, {Stefanik},
  {Esquerdo}, {Everett}, {Perumpilly}, {L{\'a}z{\'a}r}, {Papp}, \&
  {S{\'a}ri}}]{hartman2011138}
{Hartman}, J.~D., {Bakos}, G.~{\'A}., {Kipping}, D.~M., {et~al.}
  2011{\natexlab{a}}, \apj, 728, 138, 138

\bibitem[{{Hartman} {et~al.}(2011{\natexlab{b}}){Hartman}, {Bakos}, {Torres},
  {Latham}, {Kov{\'a}cs}, {B{\'e}ky}, {Quinn}, {Mazeh}, {Shporer}, {Marcy},
  {Howard}, {Fischer}, {Johnson}, {Esquerdo}, {Noyes}, {Sasselov}, {Stefanik},
  {Fernandez}, {Szklen{\'a}r}, {L{\'a}z{\'a}r}, {Papp}, \&
  {S{\'a}ri}}]{hartman201159}
{Hartman}, J.~D., {Bakos}, G.~{\'A}., {Torres}, G., {et~al.}
  2011{\natexlab{b}}, \apj, 742, 59, 59

\bibitem[{{Horedt}(2004)}]{horedt2004306}
{Horedt}, G.~P., ed. 2004, Astrophysics and Space Science Library, Vol. 306,
  {Polytropes - Applications in Astrophysics and Related Fields}

\bibitem[{{Hubbard}(1974)}]{hubbard197442}
{Hubbard}, W.~B. 1974, Icarus, 23, 42, 42

\bibitem[{{Jackson} {et~al.}(2008){Jackson}, {Greenberg}, \&
  {Barnes}}]{jackson20081396}
{Jackson}, B., {Greenberg}, R., \& {Barnes}, R. 2008, \apj, 678, 1396, 1396

\bibitem[{{Ketchum} {et~al.}(2013){Ketchum}, {Adams}, \&
  {Bloch}}]{ketchum201371}
{Ketchum}, J.~A., {Adams}, F.~C., \& {Bloch}, A.~M. 2013, \apj, 762, 71, 71

\bibitem[{{Laskar} {et~al.}(2012){Laskar}, {Bou{\'e}}, \&
  {Correia}}]{laskar2012105}
{Laskar}, J., {Bou{\'e}}, G., \& {Correia}, A.~C.~M. 2012, \aap, 538, A105,
  A105

\bibitem[{{Lee}(2004)}]{lee2004517}
{Lee}, M.~H. 2004, \apj, 611, 517, 517

\bibitem[{{Maness} {et~al.}(2007){Maness}, {Marcy}, {Ford}, {Hauschildt},
  {Shreve}, {Basri}, {Butler}, \& {Vogt}}]{maness200790}
{Maness}, H.~L., {Marcy}, G.~W., {Ford}, E.~B., {et~al.} 2007, \pasp, 119, 90,
  90

\bibitem[{{Mardling}(2007)}]{mardling20071768}
{Mardling}, R.~A. 2007, \mnras, 382, 1768, 1768

\bibitem[{{Matsumura} {et~al.}(2010){Matsumura}, {Peale}, \&
  {Rasio}}]{matsumura20101995}
{Matsumura}, S., {Peale}, S.~J., \& {Rasio}, F.~A. 2010, \apj, 725, 1995, 1995

\bibitem[{{Matsumura} {et~al.}(2008){Matsumura}, {Takeda}, \&
  {Rasio}}]{matsumura200829}
{Matsumura}, S., {Takeda}, G., \& {Rasio}, F.~A. 2008, \apjl, 686, L29, L29

\bibitem[{{Motz}(1952)}]{motz1952562}
{Motz}, L. 1952, \apj, 115, 562, 562

\bibitem[{{Murray} \& {Dermott}(1999)}]{murray1999ssd}
{Murray}, C.~D., \& {Dermott}, S.~F. 1999

\bibitem[{{Namouni}(2007)}]{namouni2007233}
{Namouni}, F. 2007, in Lecture Notes in Physics, Berlin Springer Verlag, Vol.
  729, Lecture Notes in Physics, Berlin Springer Verlag, ed. {D.~Benest,
  C.~Froeschle, \& E.~Lega}, 233

\bibitem[{{Peale}(1986)}]{peale1986159}
{Peale}, S.~J. 1986, in Satellites, ed. J.~A. {Burns} \& M.~S. {Matthews}
  (Tuscon, AZ, USA: Univ. of Arizona Press), 159--223

\bibitem[{{Pont} {et~al.}(2011){Pont}, {Husnoo}, {Mazeh}, \&
  {Fabrycky}}]{pont20111278}
{Pont}, F., {Husnoo}, N., {Mazeh}, T., \& {Fabrycky}, D. 2011, \mnras, 414,
  1278, 1278

\bibitem[{{Ragozzine} \& {Wolf}(2009)}]{ragozzine20091778}
{Ragozzine}, D., \& {Wolf}, A.~S. 2009, \apj, 698, 1778, 1778

\bibitem[{{Rasio} {et~al.}(1996){Rasio}, {Tout}, {Lubow}, \&
  {Livio}}]{rasio19961187}
{Rasio}, F.~A., {Tout}, C.~A., {Lubow}, S.~H., \& {Livio}, M. 1996, \apj, 470,
  1187, 1187

\bibitem[{{Rauch} \& {Hamilton}(2002)}]{rauch2002dda}
{Rauch}, K.~P., \& {Hamilton}, D.~P. 2002, in Bulletin of the American
  Astronomical Society, Vol.~34, Bull. Am. Astron. Soc., 938

\bibitem[{{Shen} \& {Turner}(2008)}]{shen2008553}
{Shen}, Y., \& {Turner}, E.~L. 2008, \apj, 685, 553, 553

\bibitem[{{Smalley} {et~al.}(2011){Smalley}, {Anderson}, {Collier Cameron},
  {Hellier}, {Lendl}, {Maxted}, {Queloz}, {Triaud}, {West}, {Bentley}, {Enoch},
  {Gillon}, {Lister}, {Pepe}, {Pollacco}, {Segransan}, {Smith}, {Southworth},
  {Udry}, {Wheatley}, {Wood}, \& {Bento}}]{smalley2011130}
{Smalley}, B., {Anderson}, D.~R., {Collier Cameron}, A., {et~al.} 2011, \aap,
  526, A130, A130

\bibitem[{{Winn} {et~al.}(2010){Winn}, {Johnson}, {Howard}, {Marcy}, {Bakos},
  {Hartman}, {Torres}, {Albrecht}, \& {Narita}}]{winn2010575}
{Winn}, J.~N., {Johnson}, J.~A., {Howard}, A.~W., {et~al.} 2010, \apj, 718,
  575, 575

\bibitem[{{Wright} {et~al.}(2009){Wright}, {Upadhyay}, {Marcy}, {Fischer},
  {Ford}, \& {Johnson}}]{wright20091084}
{Wright}, J.~T., {Upadhyay}, S., {Marcy}, G.~W., {et~al.} 2009, \apj, 693,
  1084, 1084

\bibitem[{{Wu} \& {Goldreich}(2002)}]{wu20021024}
{Wu}, Y., \& {Goldreich}, P. 2002, \apj, 564, 1024, 1024

\bibitem[{{Zakamska} {et~al.}(2011){Zakamska}, {Pan}, \&
  {Ford}}]{zakamska20111895}
{Zakamska}, N.~L., {Pan}, M., \& {Ford}, E.~B. 2011, \mnras, 410, 1895, 1895

\bibitem[{{Zhang}(2007)}]{zhang20071}
{Zhang}, K. 2007, PhD thesis,

\bibitem[{{Zhang} \& {Hamilton}(2003)}]{zhang20031485}
{Zhang}, K., \& {Hamilton}, D.~P. 2003, in Bulletin of the American
  Astronomical Society, Vol.~35, AAS/Division for Planetary Sciences Meeting
  Abstracts \#35, 1485

\bibitem[{{Zhang} \& {Hamilton}(2007)}]{zhang2007386}
{Zhang}, K., \& {Hamilton}, D.~P. 2007, Icarus, 188, 386, 386

\bibitem[{{Zhang} \& {Hamilton}(2008)}]{zhang2008267}
---. 2008, Icarus, 193, 267, 267

\end{thebibliography}

\end{document}